\documentclass[useAMS,usenatbib]{mn2e}

 \usepackage{times}

\usepackage{graphicx}

\def\squig{$\sim\!\!$}
\def\lesssim{\mathrel{\hbox{\rlap{\hbox{\lower4pt\hbox{$\sim$}}}\hbox{$<$}}}}
\def\gtrsim{\mathrel{\hbox{\rlap{\hbox{\lower4pt\hbox{$\sim$}}}\hbox{$>$}}}}

\def\arcsec{\hbox{$^{\prime\prime}$}}
\def\arcmin{$^{\prime}$}
\def\deg{\hbox{$^\circ$}}
\def\power{WHz$^{-1}$sr$^{-1}$}

\title[II. FRI classification \& space density evolution]{A sample of
  mJy radio sources at 1.4 GHz in the Lynx 
  and Hercules fields - II. Cosmic evolution of the space density of FRI radio sources}
\author[E. E. Rigby, P. N. Best and
  I. A. Snellen]{E. E. Rigby$^{1,2}$\thanks{E-mail:Emma.Rigby@astro.cf.ac.uk},  
P. N. Best$^{2}$  I. A. G. Snellen$^{3}$ \\
$^{1}$ School of Physics and Astronomy, Cardiff University, 5 The Parade, Cardiff CF24 3YB \\
$^{2}$SUPA\thanks{Scottish Universities Physics Alliance}, University
  of Edinburgh, Institute for Astronomy, Royal Observatory, Edinburgh
  EH9 3HJ, UK\\  
$^{3}$Leiden Observatory, Leiden University, Niels Bohrweg 2, NL-2300RA Leiden, The Netherlands}

\begin{document}


\pagerange{\pageref{firstpage}--\pageref{lastpage}} \pubyear{2007}

\maketitle

\label{firstpage}

\begin{abstract}
In this paper the cosmic evolution of the space density of Fanaroff \&
Riley Class I (FRI) radio sources is investigated out to $z \sim 1$,
in order to understand the origin of the differences between these and
the more powerful FRIIs. High resolution radio images are presented
of the best high redshift FRI candidate galaxies, drawn from two
fields of the Leiden Berkeley Deep Survey, and previously defined in
\citet[][Paper I]{me1}. Together with lower resolution radio
observations (both previously published in Paper I and, for a subset
of sources, also presented here) these are used to morphologically
classify the sample. Sources which are clearly resolved are classified by
morphology alone, whereas barely or unresolved sources were classified
using a combination of morphology and flux density loss in the higher
resolution data, indicative of resolved out extended 
emission. The space densities of the FRIs are then calculated as a
function of redshift, and compared to both measurements of the local
value and the behaviour of the more powerful FRIIs. The space density
of FRI radio sources with luminosities (at 1.4 GHz) $> 10^{25}$ W/Hz is enhanced by
a factor of 5--9 by $z \sim 1$, implying moderately strong evolution
of this population; this enhancement is in good agreement with models
of FRII evolution at the same luminosity. There are also indications
that the evolution is luminosity dependent, with the lower powered 
sources evolving less strongly.
\end{abstract}


\section{Introduction}

It has been known for some time that powerful ($P_{\rm
  178MHz}$ $>10^{24-25}$ \power) radio galaxies undergo strong
cosmic evolution, with density enhancements, over the local value, ranging
from $\sim 10$, to 100--1000 for the most luminous sources
at redshifts 1--2 (e.g. Wall et al. \citeyearpar{Wall80}). These results were supported
by Dunlop \& Peacock \citeyearpar[][hereafter DP90]{dp90} who found
that including a high luminosity, strongly evolving population in
their modelling of the radio luminosity function fitted the data
well. 

The increased difficulty in detecting lower powered radio galaxies over comparable
distances means that their evolutionary behaviour is
less clear. Initial, low redshift studies, suggested that they had a constant space density (e.g. Jackson \& Wall
\citeyearpar{Jack}) and this was subsequently supported at higher redshift by Clewley \& Jarvis
\citeyearpar{Clewley}. However, DP90 found that their models were
better fit if the low luminosity population was weakly evolving, and
Sadler et al. \citeyearpar{Sadler}, using a sample selected in
a similar way to that of Clewley \& Jarvis, found that the
no--evolution scenario is ruled out at a significance level of $> 6
\sigma$. Additionally, Willott et al. \citeyearpar{Willott}, also
see evolution in their low luminosity, weak emission line,
population. 

This high/low power radio galaxy split corresponds to the division
between Fanaroff \& Riley Class I and II sources (FRI and FRII; Fanaroff \& Riley,
\citeyear{FR}). Of the two types, FRIIs are the more powerful, but it should be noted that there is significant overlap at
the break luminosity. The bulk of the emission in FRIIs originates
from the hotspots at the ends of their highly collimated,
`edge--brightened', jets. FRIs, on the other hand, typically have
`edge--darkened' jets which are brighter towards their central regions
and which also tend to flare out close to the nucleus. It is not yet
clear whether the differences between the two classes arise
from fundamental differences in their central engines
(e.g. advection dominated accretion flow in FRIs arising from a lower
accretion rate) as argued by e.g. \citet{Zirbel2}, or whether it is the
environments in which the jets of these objects reside which determines their
observed properties (e.g. Gopal--Krishna \& Wiita
\citeyearpar{Gopal} and Gawro\'{n}ski et al. \citeyearpar{Gawronski}),
or even some combination of the two. 

An important tool, therefore, for investigating the morphological differences between the two FR
classes comes from measurements of their cosmic evolution. Since the FRI/FRII break luminosity is
not fixed, but is a function of host galaxy magnitude \citep{Ledlow},
morphological classification of the radio galaxies in a sample is a
necessity if an accurate picture of the cosmic evolution of FRIs is to be obtained. Any
similarities detected in the variation of FRIs and FRIIs over cosmic 
time lends weight to the extrinsic, environmental, model since objects
with similar underlying properties, at the same luminosity, would be
expected to evolve similarly as well. If this is the case, then it
implies that FRIs and FRIIs may simply represent different stages in
the evolution of a radio galaxy, which, in this scenario, would start
as a high luminosity FRII with powerful jets, which then weakens and
dims over time (e.g. Willott et al. \citeyearpar{Willott}; Kaiser \&
Best \citeyearpar{Kaiser}).  

In addition to clarifying the FRI/II differences, the cosmic evolution
of FRIs in particular may also be important for galaxy formation and evolution
models, which are increasingly using radio galaxies to provide the
feedback necessary to halt the problem of massive galaxy overgrowth
\citep[e.g.][]{Bower}. Best et al., \citeyearpar{Best} suggest that it
is the lower luminosity sources, perhaps the FRI population alone,
that are mainly responsible for this. Understanding their behaviour
may therefore provide the key to understanding this mechanism as
well. 

Snellen \& Best \citeyearpar{Ignas} made a first attempt at
determining the space density of morphologically classified FRIs,
using the two $z > 1$ FRIs detected in the Hubble Deep and Flanking
Fields (HDF+FF) area. The small number of sources meant that a direct
calculation would have been unreliable, but they were able to show that the
probability of detecting these two FRIs if the population undergoes no
evolution was $< 1$\%, and was instead consistent with a space density
enhancement comparable to that of less luminous FRII galaxies at the
same redshift. More recently, Jamrozy \citeyearpar{Jamrozy}, using the
number counts of two complete, morphologically classified, radio samples, also found that a
positive cosmic evolution for the most luminous FRI sources is needed
to fit their observational data out to $z \sim 2$. 

The aim of this work is to improve on the results of these previous analyses using a deep, wide field, 1.4
GHz, Very Large Array (VLA) A--array survey, an order of magnitude
larger than the HDF+FF; this area is large enough 
to allow the space density to be directly measured for the first
time. The survey was split over two fields of 0.29 sq. degrees each -- one in the constellation
of Lynx at right ascension, $\alpha=8h45$, declination,
$\delta=+44.6\deg$ (J2000) and one in Hercules at $\alpha=17h20$,
$\delta=+49.9\deg$ (J2000) -- which were originally observed as part of the
Leiden--Berkeley Deep Survey (LBDS; Kron, \citeyear{kron_80}; Koo \&
Kron, \citeyear{koo}). These fields are ideal for this work due to the
existence of previous low resolution (12.5\arcsec) radio observations
(Windhorst et al. \citeyear{Windhorst}; Oort \& Windhorst
\citeyear{oort_85}; Oort \& van Langevelde \citeyear{oort}). These can
be used, in conjunction with higher resolution data, to look for flux
density losses in the sample, which in turn can be used for FRI
classification (see \S\ref{class_sec} for full details). Alongside
this, the Hercules field has some previous optical and spectroscopic
observations by Waddington et al., \citeyearpar{Waddington}, whilst
the Lynx field is also covered by the Sloan Digital Sky Survey (York
et al. \citeyear{york}; Stoughton et al. \citeyear{stoughton}).

The survey consists of a complete sample of 81 radio sources, evenly
spread over the two fields, all above the limiting flux density,
$S_{\rm 1.4 GHz}$, of 0.5 mJy in the A--array radio data. Optical and
infra--red observations with the Isaac Newton Telescope (INT) and the
UK Infra--red Telescope (UKIRT) respectively, resulted in a
host--galaxy identification fraction of 85\%, with 12 sources
remaining unidentified at a level of
\emph{r$^{\prime}$}$\geqslant$25.2 mag (Hercules; 4 sources) or
\emph{r$^{\prime}$}$\geqslant$24.4 mag (Lynx; 7 sources) or
\emph{K}$\gtrsim$20 mag. New spectroscopic data obtained with the
Telescopio Nazionale Galileo (TNG), combined with previously published
results meant that the redshift completeness was 49\%. The redshifts
for the remaining sources with a host galaxy detection were estimated
using either the K--z or r--z magnitude--redshift relations (Willott
et al. \citeyear{WillottKz}; Snellen et al. \citeyear{SnellenRZ}). Full details of the observations used to define
the sample, and the resulting 1.4 GHz flux densities, magnitudes and
redshifts, can be found in Rigby, Snellen \& Best \citeyearpar[][hereafter Paper I]{me1}.  

In this paper we present the additional radio observations, both low
and high resolution, along with the FRI source classification and
space density measurements. The layout of the paper is as follows:
Section 2 describes the low resolution VLA B--array observations of
the Lynx field; Section 3 then describes the high redshift FRI
candidate selection and their high resolution radio follow--up. 
Section 4 outlines the method used to classify the sample and Section
5 then describes the steps taken to measure the space density of the
identified FRIs. Finally Section 6 discusses the results and
presents the conclusions of this work. Throughout this paper values for
the cosmological parameters of H$_{0} = 71$~km~s$^{-1}$Mpc$^{-1}$,
$\Omega_{\rm m} = 0.27$ and $\Omega_{\rm \Lambda} = 0.73$ are used.  

\section{Low resolution Lynx field radio observations}

The aim of the low--resolution Lynx field VLA B--array observations
was to provide a measure of the total flux density for each
field source, and to look for any extended emission that may have been
resolved out in the 1.5\arcsec\ resolution A--array data presented in
Paper I. Complimentary observations of the Hercules field in this
configuration were not done due to time constraints.   

The observations took place on 30th October 2003, in L--band (1.4 GHz),
using the wide--field mode. The flux, point and phase calibrators used were 3C286, 3C147 and 0828+493 respectively and the total
exposure time on the field was 7000s; this was split into five periods
which were interspersed with $\sim$120s visits to the phase
calibrator. 8 channels, at frequencies of 1.474 and 1.391 GHz, with a
total bandwidth of 25~MHz, were used, to minimise bandwidth smearing
effects. Full polarization was observed. 

The data were calibrated using the NRAO AIPS package. 
Since sources located at all positions in the final radio image need
to be considered, the non--coplanar array wide--field imaging
techniques, incorporated into the AIPS task IMAGR, were
used. The field was split into facets, centred on 
the positions of each source from the A--array observations and the
centre was shifted to the centre of each facet and imaged and
deconvolved in turn. Each facet was 256 by 256 pixels (with 1.0\arcsec\ per pixel).
 
The deconvolution and CLEANing of each facet was done using 5000 iterations and nearly natural
weighting to minimise the noise in the resulting CLEAN image. No
self--calibration was carried out due to the weakness of the
sources. The final noise level reached for all the facets was
$\sim$50 $\umu$Jy/beam, with a resolution of 5.36\arcsec\ by
4.67\arcsec. Finally a primary beam correction was applied to each image
to account for the attenuation of the beam away from the pointing
centre.

\subsection{Source detection and flux density measurement}

All the sources in the Lynx field complete sample were detected in
these observations. Their flux densities were, as with the A--array
data (see Paper I), measured with \emph{tvstat} if they showed extension or with
\emph{imfit} if they appeared compact. The method used for each
source in the complete sample, along with the resulting flux densities and primary beam
correction factors, can be found in Table \ref{b_fluxes}. The
corresponding radio contour maps (without primary--beam corrections applied) can be
found in Figure \ref{b_images}. The results for additional sources not in the
complete sample can be found in Table \ref{b_fluxes_no} and Figure
\ref{rad_b_im_no}; these sources were excluded as they did not fall
within the field of view of the optical imaging described in Paper I. 

\begin{table}
\centering
\caption{\protect\label{b_fluxes} The primary--beam corrected
  flux densities for the (complete sample) Lynx field B--array observations along with the
  correction factors, $C_{\rm PB}$, used. An \emph{I} in the final column
  indicates an \emph{imfit} measured flux density; a \emph{T}
  indicates a \emph{tvstat} measurement. A primary beam correction
  error of 20\% of the difference between the corrected and
  un--corrected flux density has been incorporated into the quoted
  errors. }
\begin{tabular}{l|c|c|c}
\hline
Name & S$_{\rm 1.4GHz}$ (mJy) & $C_{\rm PB}$ & Measure   \\
\hline
55w116  & 2.05 $\pm$ 0.43 & 2.22 &T \\
55w118  & 0.85 $\pm$ 0.13 & 1.92 &I \\
55w120  & 1.91 $\pm$ 0.43 & 2.68 &T \\
55w121  & 1.24 $\pm$ 0.14 & 1.60 &I \\
55w122  & 0.76 $\pm$ 0.18 & 1.45 &I \\
55w123  & 1.09 $\pm$ 0.14 & 1.33 &I \\
55w124  & 2.58 $\pm$ 0.10 & 1.35 &I \\
55w127  & 1.72 $\pm$ 0.11 & 1.36 &I \\
55w128  & 4.10 $\pm$ 0.56 & 2.05 &T \\
55w131  & 1.20 $\pm$ 0.20 & 1.48 &T \\
55w132  & 1.83 $\pm$ 0.36 & 2.05 &T \\
55w133  & 2.17 $\pm$ 0.13 & 1.47 &I \\
55w135  & 2.61 $\pm$ 0.37 & 1.98 &T \\
55w136  & 0.90 $\pm$ 0.11 & 1.10 &I \\
55w137  & 1.70 $\pm$ 0.17 & 1.29 &T \\
55w138  & 1.81 $\pm$ 0.12 & 1.37 &I \\
55w140  & 0.58 $\pm$ 0.09 & 1.06 &I \\
55w141  & 0.60 $\pm$ 0.60 & 1.19 &I \\
55w143a & 2.22 $\pm$ 0.11 & 1.34 &I \\
55w143b & 0.65 $\pm$ 0.17 & 1.33 &I \\
55w147  & 2.24 $\pm$ 0.14 & 1.82 &I \\
55w149  & 7.63 $\pm$ 1.14 & 3.20 &T \\
55w150  & 0.60 $\pm$ 0.14 & 1.88 &I \\
55w154  & 12.90$\pm$ 0.44 & 1.13 &T \\
55w155  & 1.63 $\pm$ 0.15 & 1.55 &I \\
55w156  & 4.02 $\pm$ 0.26 & 1.19 &T \\
55w157  & 1.61 $\pm$ 0.13 & 1.68 &I \\
55w159a & 6.61 $\pm$ 0.21 & 2.69 &I \\
55w159b & 1.08 $\pm$ 0.31 & 2.55 &I \\
55w160  & 0.77 $\pm$ 0.09 & 1.32 &I \\
55w161  & 0.87 $\pm$ 0.16 & 1.87 &I \\
55w165a &17.51 $\pm$ 1.33 & 2.06 &T \\
55w165b & 1.40 $\pm$ 0.23 & 1.99 &I \\
55w166  & 2.26 $\pm$ 0.17 & 2.07 &I \\
60w016  & 0.57 $\pm$ 0.15 & 1.52 &I \\
60w024  & 0.25 $\pm$ 0.08 & 1.29 &I \\
60w032  & 0.29 $\pm$ 0.10 & 1.51 &I \\
60w039  & 0.64 $\pm$ 0.15 & 1.38 &I \\
60w055  & 0.80 $\pm$ 0.27 & 2.34 &I \\
60w067  & 0.65 $\pm$ 0.11 & 1.70 &I \\
60w071  & 0.59 $\pm$ 0.19 & 1.42 &I \\
60w084  & 0.54 $\pm$ 0.16 & 2.08 &I \\
\hline
\end{tabular}
\end{table}

\begin{table}
\centering
\caption{The primary--beam corrected
  flux densities for the sources not included in the Lynx complete
  sample, along with the correction factors, $C_{\rm PB}$, used. An \emph{I} in the final column
  indicates an \emph{imfit} measured flux density; a \emph{T}
  indicates a \emph{tvstat} measurement. A primary beam correction
  error of 20\% of the difference between the corrected and
  un--corrected flux density has been incorporated into the quoted
  errors.\label{b_fluxes_no}}
\begin{tabular}{l|c|c|c}
\hline
Name & S$_{\rm 1.4GHz}$ (mJy) & $C_{\rm PB}$ & Measure   \\
\hline
55w119 & 1.97 $\pm$ 0.36 & 3.86& I     \\
55w125  &18.87$\pm$ 4.04 &12.01& I     \\
55w126  & 4.13$\pm$ 0.90 &6.37 & I     \\
\hline
\end{tabular}
\end{table}

\section{FRI candidate selection and high resolution radio observations}

The high--resolution radio observations were, by necessity, limited to
the best high--redshift FRI candidates only. The next step therefore,
was to select these candidate sources from the sample. The candidate
criteria were, firstly, that their redshifts (or estimated redshifts) were $\gtrsim 1.0$ 
and, secondly, that extended emission was visible in their radio
image; this extension was defined, by inspection, as structure that deviated
from a compact form. Since the high--resolution observations took
place using two different instruments, the VLA+PieTown (VLA+Pt) link
(\S\ref{pietown}) and the MERLIN (Multi Element Radio Linked Interferometer) array
(\S\ref{merlin}), with slightly different characteristics, as
described below, two different subsamples of these candidates were observed. Where possible, candidate
sources were included in both observations to increase the possibility
of detecting and classifying significant extended emission. Three of the high--redshift
candidates were not able to be included in either observation due to
time constraints; these were 53w081 in Hercules and 55w155 and 55w166 in Lynx.  

\subsection{VLA A+PieTown observations \protect\label{pietown}}

The 27 radio antennae of the VLA can be linked with one of the
antennas of the Very Long Baseline Array (VLBA), located 50 km
away from the array centre at Pie Town. The extra baselines this adds
means that the VLA+Pt can reach sub--arcsecond resolution at 1.4 GHz. 

The candidate sources in the Hercules field, listed in Table
\ref{pt_obs}, were individually observed on 26th September 2004 with the VLA+Pt at
1.4 GHz (L band). The candidate Lynx sources were observed
similarly on 18th February 2006. 50 MHz channels at 1.385 and 1.465 GHz were used for both observations. The
details of the exposure times for each source can be found in Table
\ref{pt_obs}. The flux and point calibrator calibrator for both fields was 3C286 and the
phase calibrators were 1727+455 for the Hercules and 0832+492 for the Lynx field. 

\begin{table}
\centering
\caption{Details of the VLA+Pt exposure times for the Lynx and Hercules
  high--redshift candidate sources, along with the resulting flux
  densities and noise limits. \protect\label{pt_obs}\protect\label{rad_pt_table} }
\begin{tabular}{c|c|c|c}
\hline
\multicolumn{4}{l}{Hercules} \\
\hline
Source Name & Exposure Time (s) & $S_{\rm 1.4 GHz}$ (mJy) & rms ($\umu$Jy) \\ 
\hline
53w054a & 2600 & 1.82  $\pm$ 0.07 & 29 \\   
53w054b & 2600 & 2.58  $\pm$ 0.07 & 29 \\   
53w059  & 2580 & 21.23 $\pm$ 0.21 & 30 \\   
53w061  & 2580 & 1.68  $\pm$ 0.06 & 28 \\   
53w065  & 2590 & 6.26  $\pm$ 0.09 & 33 \\   
53w069  & 2570 & 3.50  $\pm$ 0.18 & 36 \\   
53w087  & 2600 & 4.19  $\pm$ 0.17 & 28 \\   
53w088  & 2590 & 14.11 $\pm$ 0.09 & 30 \\   
\hline
\multicolumn{4}{l}{Lynx} \\
\hline
55w116  & 2550 & 0.91  $\pm$ 0.16 & 49 \\   
55w120  & 2570 & 0.98  $\pm$ 0.13 & 47 \\   
55w121  & 2560 & 1.32  $\pm$ 0.11 & 43 \\   
55w128  & 2570 & 2.94  $\pm$ 0.45 & 47 \\   
55w132  & 2560 & 1.02  $\pm$ 0.19 & 45 \\   
55w133  & 2570 & 2.23  $\pm$ 0.14 & 44 \\   
55w136  & 2580 & 0.76  $\pm$ 0.17 & 44 \\   
55w138  & 2570 & 1.74  $\pm$ 0.18 & 45 \\   
\hline
\end{tabular}
\end{table}

The data were again calibrated using the NRAO AIPS package. Each
calibrated source observation was edited to remove bad data and then
CLEANed using 10000 iterations, with a pixel size of 0.15\arcsec\, and
nearly pure uniform weighting to avoid downweighting the longest, PieTown, baselines and thus achieve
high resolution images. The noise level reached for each source
observation is given in Table \ref{rad_pt_table} above and the
resolution of the images was typically 1.1\arcsec$\times$0.6\arcsec\ 
with the highest resolution direction determined by the location of
the PieTown antenna.

All the candidate sources included in the VLA+Pt observations were
detected and their flux densities were then measured with
\emph{tvstat}; these can be found in Table \ref{rad_pt_table}. The
corresponding contour plots can be found in Figures 
\ref{comp_im_h} and \ref{comp_im}.

\subsection{MERLIN observations \protect\label{merlin}}

The MERLIN observations were done in wide--field mode, with an
individual field size of about 2.5\arcmin\ radius;
at this distance from the pointing centre the bandwidth smearing is
$<10$\%. Three of these sub--fields were used in the Hercules field
and 4 in the Lynx field. The pointing positions were designed such that
the maximum number of non--candidate sources could also be observed whilst still including
as many of the candidates as possible. Whilst these extra sources were not selected as high--redshift FRI
candidates, observing them at higher--resolution may help to classify
some of the lower redshift objects in the sample. 53w091, one of the sources not included in the complete sample, 
was also included in the MERLIN observations as a result of this
process. 

MERLIN consists of up to seven antennas spread across the UK, giving
baselines up to 217 km and a resolution, at L--band, of
$\sim$0.15\arcsec. The sources were observed with MERLIN on 11th March 2005 for the Hercules and on 13th,
19th, 20th, 28th and 29th May 2006 for the Lynx field, both
using 32 channels, with a combined bandwidth of 15.5 MHz. It should be noted that due to technical problems the Lovell
telescope, the largest antenna in the array, was only included in the
observations of Lynx B, C and D; this
increased the sensitivity for these sub--fields by a factor of
\squig\ 2.2. 

For the Hercules subfields, the flux calibrators used were 3C286 and
OQ208 and the phase calibrator was 1734+508. For three of the Lynx
subfields (B, C and D), the flux calibrators were again 3C286 and
OQ208; for Lynx A, however, 3C286 and 2134+004 were used. The Lynx phase
calibrator was 0843+463. Two flux calibrators are necessary as 3C286
is resolved on all but the shortest MERLIN baselines. All subfields
were observed for 12 hours. 

The initial editing and calibration of the data was done at Jodrell
Bank using local software specially written for the MERLIN
array. Following this the data were loaded into AIPS and the MERLIN
pipeline \citep{MUG}, which uses the AIPS calibration tasks with
MERLIN specific inputs, was then used to complete the
calibration. Data which included the Lovell telescope were then
reweighted to account for the different sensitivities of the 
instruments in the array. 

Finally each subfield was Fourier transformed and deconvolved to form
the final CLEANed image. To minimise the non--coplanar array effects
the centre of each subfield was again shifted to the position of each
source it contained and treated separately. Each source facet was 512 by 512 pixels with
0.045\arcsec/pixel. The rms noise reached for each source was between 27 and 49 $\umu$Jy, and can be found
in Tables \ref{merlin_flx_h} and \ref{merlin_flx_l}. This process was then repeated for
Lynx subfields B, C and D, with all baselines including the Lovell
telescope removed from the data. This was done as, without Lovell, a 76 m diameter dish, the MERLIN antennae are all comparable in size
(similar to those of the VLA) and so a simple primary beam correction method can be applied. The
non--Lovell images were therefore used to measure the primary--beam
corrected flux densities of the sources but the deeper,
Lovell included, images were used for the morphological
classification. The rms noise reached in the non--Lovell images can
also be found in Table \ref{merlin_flx_l}.

For the Hercules sources, one candidate and one of the `extra' sources
were resolved out; the rest were all detected. In the Lynx subfields
all the candidates were detected and 3 of the `extra' sources were
resolved out. A primary beam correction was then applied to correct
for the attenuation of the primary beam on the off--centre sources and
the flux densities were measured; the resulting values can be found in Tables \ref{merlin_flx_h} and
\ref{merlin_flx_l}. The corresponding contour maps can be found in Figures \ref{comp_im_h}
and \ref{comp_im}. 

\begin{table}
\centering
\caption{The primary beam corrected flux densities and
  un--primary beam corrected noise levels found from the Hercules MERLIN observations. A
  * indicates a source which was resolved out. All flux densities were
  measured using \emph{tvstat}. A primary beam correction
  error of 20\% of the difference between the corrected and
  un--corrected flux density has been incorporated into the quoted
  errors.\protect\label{merlin_flx_h}} 
\begin{tabular}{c|c|c|c|c}
\hline
\multicolumn{5}{l}{Hercules} \\
\hline
Sub--field & Name & $S_{\rm 1.4 GHz}$ (mJy)& $C_{\rm PB}$ &
rms ($\umu$Jy)\\
\hline
A & 53w054a & 1.47  $\pm$ 0.13& 1.03 & 41\\  
A & 53w054b & 1.83  $\pm$ 0.13& 1.02 & 41\\
A & 53w057  & 2.04  $\pm$ 0.14& 1.00 & 40\\
B & 53w059  & 18.87 $\pm$ 0.62& 1.06 & 46\\
A & 53w061  & 1.06  $\pm$ 0.10& 1.03 & 34\\
B & 53w065  & 4.62  $\pm$ 0.18& 1.05 & 46\\
B & 53w066  & 3.32  $\pm$ 0.18& 1.00 & 47\\
B & 53w070  & 2.29  $\pm$ 0.15& 1.08 & 46\\
C & 53w087  &  *              &      & 43\\
C & 53w088  & 10.80 $\pm$ 0.18& 1.05 & 49\\ 
\hline
\multicolumn{5}{l}{Extra sources} \\
\hline
C & 53w082  & 2.40 $\pm$ 0.15 & 1.09 & 45\\
C & 53w089  &   *             &      & 42\\
C & 53w091  &32.81 $\pm$ 1.00 & 1.04 & 47\\
\hline
\end{tabular}
\end{table}

\begin{table*}
\centering
\caption{The primary beam corrected flux densities and
  un--primary beam corrected noise levels, both with and without the Lovell
  telescope in the array, found from the Lynx MERLIN observations. A
  * indicates a source which was resolved out. A primary beam correction
  error of 20\% of the difference between the corrected and
  un--corrected flux density has been incorporated into the quoted
  errors. A `T' indicates a \emph{tvstat} measurement and an `I'
  indicates an \emph{imfit} measurement. \protect\label{merlin_flx_l}} 
\begin{tabular}{c|c|c|c|c|c|c}
\hline
\multicolumn{7}{l}{Lynx} \\
\hline
Sub--field & Name & $S_{\rm 1.4 GHz}$ (mJy)& $C_{\rm PB}$ & Measure & 
rms (Lovell) & rms (no Lovell)\\
 & & & & &($\umu$Jy) & ($\umu$Jy)\\
\hline
C & 55w116 & 1.68 $\pm$ 0.34& 1.06 &T& 30& 68\\
C & 55w121 & 0.97 $\pm$ 0.14& 1.05 &I& 32& 71\\
B & 55w128 & 1.52 $\pm$ 0.32& 1.04 &T& 29& 67\\
B & 55w132 & 1.29 $\pm$ 0.31& 1.01 &T& 28& 65\\
D & 55w133 & 1.63 $\pm$ 0.17& 1.04 &I& 30& 68\\
D & 55w136 & 0.44 $\pm$ 0.19& 1.04 &T& 28& 68\\
B & 55w138 & 1.16 $\pm$ 0.26& 1.01 &T& 30& 67\\
D & 55w143a& 1.73 $\pm$ 0.20& 1.05 &T& 37& 62\\
A & 55w159a& 4.71 $\pm$ 0.18& 1.04 &I& --& 98\\
\hline
\multicolumn{6}{l}{Extra sources} \\
\hline
C & 55w118  & 0.77 $\pm$ 0.19 & 1.01 &I& 32  & 68\\
C & 55w122  & 0.32 $\pm$ 0.15 & 1.07 &I& 30  & 67\\
C & 55w123  & 1.01 $\pm$ 0.12 & 1.03 &I& 33  & 70\\
C & 55w124  & 3.26 $\pm$ 0.14 & 1.05 &I& 32  & 76\\
B & 55w127  & 1.31 $\pm$ 0.31 & 1.08 &T& 27  & 64\\
B & 55w131  & *               &      & & 28  & 65\\
B & 55w137  & 1.14 $\pm$ 0.17 & 1.02 &T& 28  & 63\\
D & 55w141  & 0.26 $\pm$ 0.10 & 1.02 &I& 32  & 66\\
D & 55w143b & 0.31 $\pm$ 0.11 & 1.04 &I& 33  & 67\\
A & 55w150  &  *              &      & & --  & 102\\
A & 55w157  & 2.51 $\pm$ 0.30 & 1.04 &I& --  & 108\\
A & 55w159b &  *             & 1.04   & & --  & 96\\
C & 60w016  & 0.48 $\pm$ 0.16 & 1.02 &T& 33  & 72\\
\hline
\end{tabular}
\end{table*}

\subsection{Comparison of results}
\label{rad_comp_here}

Comparing the flux density values at the different radio resolutions, it is clear that, for many of the sources, the
measured flux density decreases as the resolution of the observations
increases. For the non--quasar sources, it is likely
that this loss indicates the presence of resolved--out, 
extended emission, which in the absence of hotspots may indicate an
FRI--type structure. The four quasars in the sample are variable at
some level, so any flux density loss they exhibit may be due to this.

Since the resolutions of the four radio observations are so
different, it is hard to gain an overview of the structure of
individual sources in the sample from considering the respective contour maps separately. 
Therefore, Figures \ref{comp_im_h} and \ref{comp_im} show, for
the sources included in the VLA+Pt or MERLIN observations, all the
available radio contour maps at the same scale, centred on the optical host
galaxy positions, if available, or otherwise on the A--array
position\footnote{The starburst galaxy 55w127 is not included in this
  Figure as it is not an FRI candidate}. Collecting the images 
together like this illustrates the power of the 
high resolution observations in differentiating FRI objects from
FRIIs. For example, the Hercules field source 53w059 is an FRI
candidate in the A--array data but it is only in the MERLIN map,
showing its inner jet and resolved out lobes, that it can be firmly
classified as such. On the other hand, the A--array image of the Lynx
field source 55w138 does not show any clear structure and whilst the
VLA+Pt map indicates that it is extended, the MERLIN map is needed to
show the location of the jet hotspots.  

\section{FRI identification and classification}
\label{class_sec}
The most secure method of FRI classification, the detection of weak
extended emission relative to a compact core, is impossible
for the majority of the sample sources due to the lack of firm jet
detections. FRI--type, as described above, can also be inferred by comparing the source
flux densities at the low and high resolutions; a drop in flux density indicates the
presence of resolved--out, extended emission, which in the absence of
hotspots is likely to be due to an FRI. 

The classifications were therefore done by inspecting the source
morphologies where possible, or by using the flux--loss method where
not, ignoring the possibly variable quasars. However, since
the comparison between the Oort et al. \citeyearpar{oort,oort_85} and the A--array data
cannot be relied upon to determine flux density loss, as discussed
in Paper I, the only loss comparison that could be done for the Hercules field
was between the A--array and VLA--Pt or MERLIN data. As 
a result, Hercules sources which were not covered by the
higher--resolution observations, and which showed no 
obvious FRI--type jets in the A--array radio maps, could not be firmly
classified. For the Lynx field sources, the existence of the B--array
observations, which are of a similar resolution to the Oort et
al. \citeyearpar{oort_85} data, meant that these could be used for a more
internally consistent comparison with the A--array, MERLIN and VLA--Pt
data, instead.  

Considering all these factors, five classification groups were
defined for the sample. Group:
\begin{enumerate}
\item = Certain FRIs -- these clearly show typical weak, edge--darkened, FRI jets and compact cores, 
\item = Likely FRIs -- these show some morphological extension
  consistent with an FRI structure, but not enough to be definitely
  classified as FRIs, along with a flux loss of 3$\sigma$ or greater
  at higher resolution,    
\item = Possible FRIs -- either no extension is seen for these sources but they
  still lose $\geq3\sigma$ of their flux when going to higher
  resolution, or some extension consistent with an FRI structure is
  seen but little flux is lost, 
\item = Not FRIs -- this group consists of sources which either have no
  flux loss or have FRII-type morphology,
\item = Unclassifiable sources -- this group is for sources in the
  Hercules field which are compact in the A--array maps, lose no flux between the Oort et
  al. and A--array data and were not included in the higher resolution
  observations.  
 \end{enumerate}

\begin{figure*}
\centering
\includegraphics[scale=0.75]{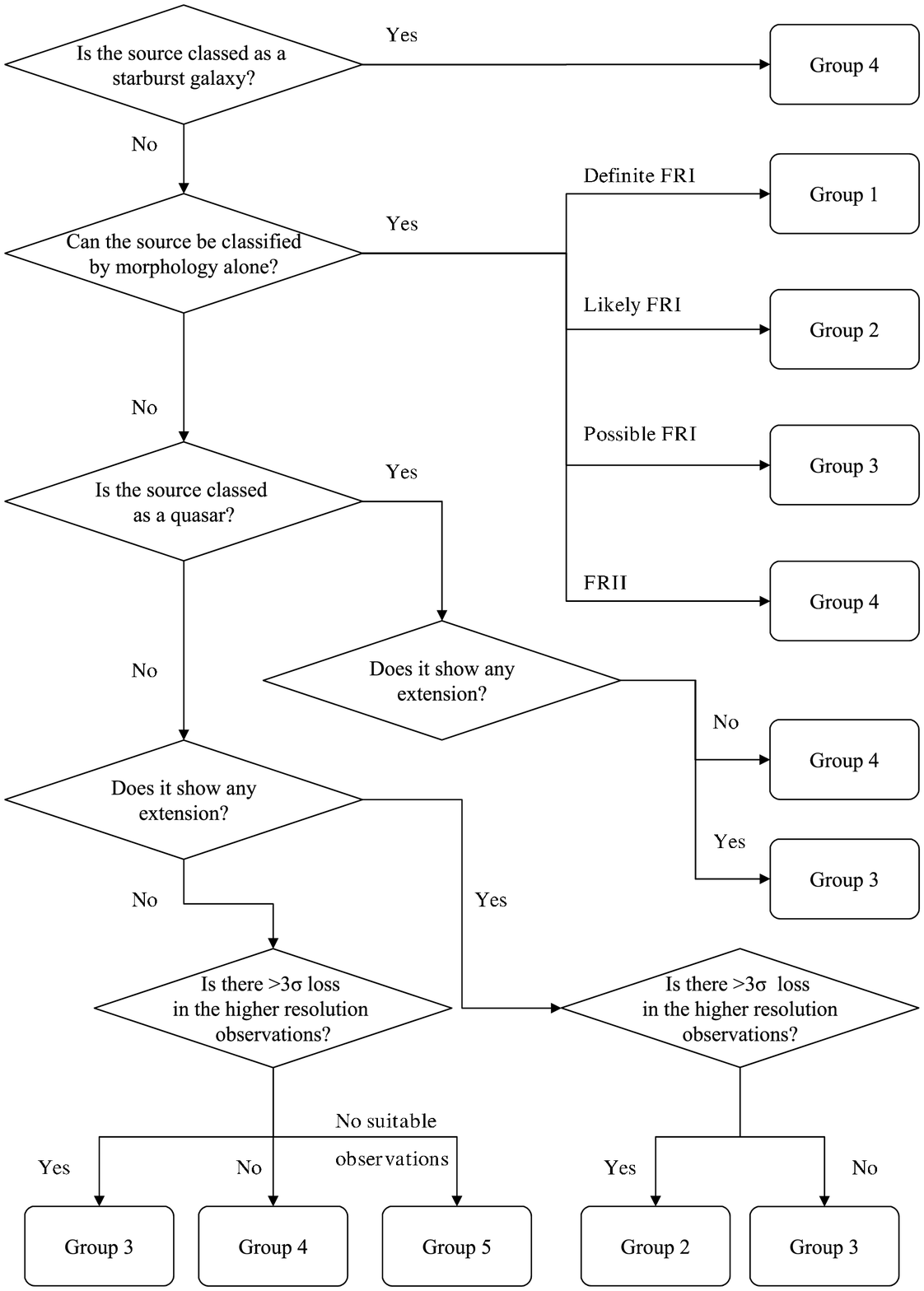}
\caption{The procedure followed for the source morphological
  classification. \protect\label{flowchart}}
\end{figure*}

The flux density loss between all the different observations was
calculated using $(1-\frac{S_{\rm hr}}{S_{\rm lr}})$, where $S_{\rm hr}$ and
$S_{\rm lr}$ are the flux densities measured in the higher and lower
resolutions respectively. The $\sigma$ values for these losses were
then found by dividing the loss by its corresponding error; these can
be found in  Tables \ref{sigma_tab_h} and \ref{sigma_tab_l} with $>$3$\sigma$
losses highlighted in bold. Each source was classified into one of the groups in turn using the
flowchart shown in Figure \ref{flowchart}. For the sources that were
classified by morphology alone, the process was repeated by
all three authors independently to ensure that the results were consistent. If
there was a disagreement between the three testers, the median grouping value
was taken. The results given by each tester, along with the final groups
assigned to these, and the rest of the sample, can be found in Tables \ref{sigma_tab_h} and
\ref{sigma_tab_l}. Additionally, the two FRIs detected in
the Hubble Deep and Flanking fields \citep{Ignas} were incorporated into the sample, and the HDF+HFF
area (10$\times$10 arcmin$^{2}$) was added in to the Hercules and Lynx areas determined by the
optical imaging. Both of these FRIs are added to group 1; since the entire
HDF+HFF area has been covered by very deep high resolution (0.2 arcsec)
radio imaging \citep{muxlow}, all sources are definitively classified
and there are no group 2 or 3 sources. The two FRI sources detected here is
slightly higher than the $\sim$1 FRI predicted for this sky area from the maximal
surface density in the Lynx and Hercules fields, but is within the errors:
including this region does not significantly change our results but does
improve the statistics.

\begin{table}
\centering
\caption{\protect\label{sigma_tab_h} The $\sigma$ flux density loss
  and classification groups (for the three testers in no particular order) for the sources in 
  the Hercules field complete sample with values of $>$3$\sigma$ 
  highlighted in bold. A, P and M represent the VLA A--array, VLA A+Pt
  and MERLIN observations respectively. A `*' indicates a source which
  was resolved out in the MERLIN observations. Sources previously
  classified as starburst galaxies and quasars are labelled with the
  superscripts `SB' and `Q' respectively.} 
\begin{tabular}{c|c|c|c|c|c|c}
\hline
Name & $\sigma_{\rm P/A}$ & $\sigma_{\rm M/A}$ & Test 1 & Test 2 & Test 3 & Final\\
\cline{4-6}
&&&\multicolumn{3}{c|}{Group}& Group\\
\hline
53w052    &          &          &  &  &  &3 \\
53w054a   &     1.06 &    2.46  &  &  &  &4 \\
53w054b   &    -0.40 &    2.22  &  &  &  &4 \\
53w057    &          &   -0.35  &  &  &  &4 \\
53w059    &     2.35 &\bf{4.33} &1 &1 &1 &1 \\
53w061$^{Q}$    &    -1.07 &    2.22  &  &  &  &3 \\
53w062    &          &          &  &  &  &5 \\
53w065    &    -2.16 &\bf{5.96} &  &  &  &2 \\
53w066    &          &\bf{5.81} &3 &3 &2 &3 \\
53w067    &          &          &1 &2 &1 &1 \\
53w069    & \bf{6.33}&          &2 &1 &1 &1 \\
53w070    &          &    1.54  &  &  &  &4 \\
53w075$^{Q}$    &          &          &  &  &  &4 \\
53w076    &          &          &2 &2 &1 &2 \\
53w077    &          &          &4 &4 &4 &4 \\
53w078    &          &          &  &  &  &3 \\
53w079    &          &          &  &  &  &5 \\
53w080$^{Q}$    &          &          &4 &4 &4 &4 \\
53w081    &          &          &  &  &  &5 \\
53w082    &          &    1.38  &  &  &  &4 \\
53w083    &          &          &  &  &  &5 \\
53w084    &          &          &  &  &  &5 \\
53w085    &          &          &  &  &  &3 \\
53w086a   &          &          &2 &3 &2 &2 \\
53w086b   &          &          &3 &3 &2 &3 \\
53w087    &\bf{15.08}&     *    &  &  &  &2 \\
53w088    &     0.14 &    1.71  &  &  &  &3 \\
53w089    &          &     *    &  &  &  &3 \\
66w009a   &          &          &  &  &  &4 \\
66w009b$^{SB}$   &          &          &  &  &  &4 \\
66w014    &          &          &  &  &  &4 \\
66w027$^{SB}$    &          &          &  &  &  &4 \\
66w031    &          &          &  &  &  &3 \\
66w035    &          &          &3 &3 &2 &3 \\
66w036    &          &          &2 &3 &1 &2 \\
66w042    &          &          &2 &2 &1 &2 \\
66w047    &          &          &  &  &  &3 \\
66w049    &          &          &  &  &  &4 \\
66w058    &          &          &  &  &  &5 \\
\hline
\end{tabular}
\end{table} 

\begin{table*}
\centering
\caption{\protect\label{sigma_tab_l} The $\sigma$ flux density loss
  and classification group (for the three testers in no particular order) for the sources in 
  the Lynx field complete sample with values of $>$3$\sigma$ 
  highlighted in bold. B, A, P and
  M represent the VLA A and B--array, VLA A+Pt
  and MERLIN observations respectively. A `*' indicates a source which
  was resolved out in the MERLIN observations. Sources previously
  classified as starburst galaxies and quasars are labelled with the
  superscripts `SB' and `Q' respectively.}  
\begin{tabular}{c|c|c|c|c|c|c|c|c|c}
\hline
Name & $\sigma_{\rm A/B}$ & $\sigma_{\rm P/B}$ & $\sigma_{\rm P/A}$ & $\sigma_{\rm M/B}$ & 
$\sigma_{\rm M/A}$ & Test 1 & Test 2 & Test 3 & Final \\
\cline{7-9}
&&&&&&\multicolumn{3}{c|}{Group}& Group\\
\hline
  55w116    &     1.30 &\bf{4.59}&    2.79 &    0.75 &   -0.36 &4 &3 &3 &3 \\
  55w118    &     0.68 &         &         &    0.36 &   -0.13 &  &  &  &4 \\
  55w120    &     0.50 &\bf{3.64}&\bf{3.26}&         &         &2 &3 &2 &2 \\
  55w121    &     1.23 &   -0.43 &   -1.54 &    1.52 &    0.40 &  &  &  &3 \\
  55w122    &     0.51 &         &         &    2.62 &    2.12 &  &  &  &3 \\
  55w123    &    -0.47 &         &         &    0.45 &    1.14 &  &  &  &4 \\
  55w124$^{Q}$    &    -1.10 &         &         &   -3.55 &   -2.01 &  &  &  &4 \\
  55w127$^{SB}$    &     0.52 &         &         &    1.26 &    1.01 &  &  &  &4 \\
  55w128    &    -0.79 &    1.92 &\bf{3.26}&\bf{6.71}&\bf{8.86}&1 &2 &1 &1 \\
  55w131    &     2.77 &         &         &    *    &    *    &2 &2 &1 &2 \\
  55w132    &     0.43 &    2.98 &    2.74 &    1.36 &    1.04 &4 &3 &3 &3 \\
  55w133    &    -0.38 &   -0.31 &    0.09 &    2.75 &\bf{3.01}&  &  &  &2 \\
  55w135$^{SB}$    &    -1.80 &         &         &         &         &  &  &  &4 \\
  55w136    &    -0.15 &    0.72 &    0.87 &    2.38 &    2.54 &4 &4 &3 &4 \\
  55w137    &     0.20 &         &         &    2.69 &    2.74 &  &  &  &3 \\
  55w138    &    -0.91 &    0.33 &    1.13 &    2.78 &\bf{3.08}&3 &4 &2 &3 \\
  55w140$^{Q}$    &     0.29 &         &         &         &         &  &  &  &4 \\
  55w141    &     1.70 &         &         &\bf{3.07}&    1.59 &  &  &  &3 \\
  55w143a   &     0.18 &         &         &    2.28 &    2.08 &3 &3 &2 &3 \\
  55w143b   & \bf{3.01}&         &         &    2.49 &    0.16 &  &  &  &3 \\
  55w147    &     1.19 &         &         &         &         &  &  &  &4 \\
  55w149    &    -0.12 &         &         &         &         &1 &1 &1 &1 \\
  55w150    &    -0.17 &         &         &         &         &  &  &  &4 \\
  55w154    &    -1.31 &         &         &         &         &1 &1 &1 &1 \\
  55w155    &    -0.33 &         &         &         &         &  &  &  &4 \\
  55w156    &    -2.01 &         &         &         &         &1 &1 &1 &1 \\
  55w157    &     2.34 &         &         &   -2.48 &   -3.26 &  &  &  &4 \\
  55w159a   &     0.14 &         &         &\bf{4.17}&    2.42 &2 &2 &2 &2 \\
  55w159b   &     0.23 &         &         &     *   &    *    &  &  &  &3 \\
  55w160    &    -0.34 &         &         &         &         &  &  &  &4 \\
  55w161    &    -1.39 &         &         &         &         &  &  &  &3 \\
  55w165a   &    -0.65 &         &         &         &         &1 &1 &1 &1 \\
  55w165b   &     2.36 &         &         &         &         &  &  &  &4 \\
  55w166    &    -0.16 &         &         &         &         &  &  &  &4 \\
  60w016    &    -1.14 &         &         &    0.45 &    2.28 &  &  &  &3 \\
  60w024    &    -0.93 &         &         &         &         &  &  &  &4 \\
  60w032    &    -0.97 &         &         &         &         &  &  &  &4 \\
  60w039$^{SB}$    &    -0.35 &         &         &         &         &  &  &  &4 \\
  60w055    &    -0.05 &         &         &         &         &  &  &  &4 \\
  60w067    &     0.74 &         &         &         &         &  &  &  &4 \\
  60w071    &     0.38 &         &         &         &         &  &  &  &4 \\
  60w084    &    -1.75 &         &         &         &         &  &  &  &3 \\
\hline
\end{tabular}
\end{table*}

\section{Investigating the FRI subsample}

Now that the radio sample has been classified, the sources in groups
1--3 (the certain, likely and possible FRIs) can be used to investigate
the changes in co--moving space density out to high redshift. To
ensure that this measurement is robust out to a significant distance,
a 1.4 GHz radio power limit ($P_{\rm lim}$) of
$\geq 10^{25}$~W/Hz, was imposed on the sample. This limit corresponds
to a maximum redshift, $z_{\rm max}$, of 1.86 out to which a source at
the 0.5~mJy sample flux density limit, and with a typical spectral index, $\alpha$, of 0.8, could be seen.

There is a possibility that group 5, the `Unclassifiable sources',
contains FRIs which should be included in the space density
analysis. However, of the 6 sources which fall into this category, half
are at $z \sim 0.6$ and half lie at $z > z_{\rm max}$. Since it is the $z
\sim 1.0$ redshift behaviour that is being investigated here, these
missing sources should not significantly affect the results.

This section describes the steps taken to measure the minimal and maximal co--moving space
density of the three FRI groups in the sample. Firstly, the parameters
and techniques used for the measurement are defined, followed by the methods used to determine the
local FRI space density, and finally, the results of the space--density calculation
are presented. 

\subsection{The evolving FRI space density}

The space densities were calculated using the $1/V_{\rm max}$
statistic, where the space density, $\rho$, of $N$ sources in some
redshift bin $\Delta z$ is simply  
\begin{equation}
\label{dens_eqn}
\rho_{\rm \Delta z} = \sum_{i=0}^{N}{\frac{w_{i}}{V_{i}}}
\end{equation}
where $V_{i}$ is the volume over which a source $i$, with weight $w_{i}$, could be seen in a
particular bin. Four redshift bins of width 0.5,
covering the range $0.0 < z \leq 2.0$ were used for the sample. The upper end of the final,
$z=1.5$ -- $2.0$, bin exceeded the value of $z_{\rm max}$ corresponding
to the limiting flux density for $\alpha = 0.8$; as a result the
individual maximum redshifts of the sources in this bin were 
determined, so that only the volumes over which each source could be
observed were used to find the density here. 

It should be noted that the redshift limit strongly depends on the
value of $\alpha$ used; as the spectral index steepens from 0.5 to 1.5, $z_{\rm max}$
decreases from 2.16 to 1.43. It is important to stress that $\alpha$ needs to be $\geq 1.37$ before
$z_{\rm max}$ falls below the start of the final redshift bin, for the
limiting power of $10^{25}$ W/Hz. This spectral index is steeper than
that found for essentially all radio sources which suggests that the
space density results in the $z < 1.5$ bins are robust to changes in
the assumed value of $\alpha$. Therefore, it is only in the final, $1.5 < z <
2.0$, bin that the values of $V_{i}$ (and hence the density) depend on
the assumed value of $\alpha$; the $V_{i}$s for sources in the other bins are all
the full bin volume. In reality though, the 3 sources that do
lie in the last bin have flux densities which are all much greater
than the 0.5 mJy limit (the faintest is 4.53 mJy), and consequently the
result does not change if $\alpha$ is varied. 

The three different FRI classification groups used meant that a
minimum (8 sources; group 1 only), maximum (16 sources; groups 1, 2
and 3) and probable (12 sources; groups 1 and 2) FRI space density
could be calculated. The flux densities used to calculate the source radio powers are those
resulting from the A--array observations (Paper I), and sources that were
previously classified as starburst galaxies have been removed from the
sample. The average luminosity in each bin can be found in Table
\ref{calc_table}; they indicate that there is a small
luminosity--redshift trend but the effect of this has been largely mitigated
by the $P_{\rm lim}$ limit. The spectral indices for 51\% of the Hercules field sources were taken
from Waddington et al. \citeyearpar{Waddington}; for the remainder of 
the sample $\alpha$ was assumed to be 0.8. The validity of this assumption was 
tested by recalculating the radio powers using two extreme $\alpha$
values of 0.5 and 1.8 for all the sources in the sample. Up to and
including these limits, the number of sources in each bin does not
change. 

\subsection{The local FRI space density}

An accurate measurement of the local FRI space density is vital for
determining whether the sample described here demonstrates a density
enhancement at high redshift. This can be obtained by directly
measuring the FRI numbers in two different local, complete, radio
samples. 

The first local measurement was carried out using a complete subsample of
the 3CR galaxy survey \citep{3cr} which contains 30 FRI sources with
$S_{\rm 178MHz}\geq 10.9$~Jy (corresponding to $S_{\rm 1.4GHz}\geq
2.09$~Jy, using $\alpha=0.8$), in an area of 4.24~sr. A source with
power $P_{\rm lim} = 10^{25}$ W/Hz and $\alpha = 0.8$, can be seen out to a redshift of 0.046 if it was
in this survey. Converting the 3CR flux densities to 1.4 GHz using
their published spectral indices \citep{3cr}, it was found that there were 4 FRIs
with $P_{\rm 1.4GHz} \geq P_{\rm lim}$ in the comoving volume of
$V({\rm z\leq 0.046})\times (4.24/4~\pi) =1.0 \times 10^{7}$~Mpc$^{3}$, which
gives a local space density of 402 $\pm$ 201 FRIs/Gpc$^{3}$.  

The second local sample used was the equatorial radio galaxy survey of
Best et al. \citeyearpar{best_eqt} which contains 178 sources, including 9
FRIs, with $S_{\rm 408MHz} > 5$~Jy (corresponding to $S_{\rm 1.4GHz} >
1.9$~Jy, again using $\alpha = 0.8$), in an area of 3.66~sr. In this
survey a source at $P_{\rm lim}$ can be seen out to $z = 0.048$; there
are 2 FRIs in this volume with powers greater or equal to this which
gives a local density of 196 $\pm$ 139 FRIs/Gpc$^{3}$.  

Combining the results from these two surveys gives a total of 6 FRI
sources in a comoving volume of 2.0 $\times$ 10$^{7}$~Mpc and a
corresponding comoving local FRI space density of 298 $\pm$ 122
FRIs/Gpc$^{3}$. 

An alternative estimate can be obtained by integrating the local radio luminosity function of Best et
al. \citeyearpar{Best_rlf}, which was calculated from a sample of 2215
radio--loud AGN, formed by comparing the SDSS with two radio surveys:
the National Radio Astronomy Observatories (NRAO) Very Large Array
(VLA) Sky Survey (NVSS) \citep{condon_nvss} and the Faint Images of the Radio Sky at
Twenty centimetres (FIRST) survey \citep{becker_first}. 

Since the radio luminosity function decreases rapidly
above $\log P = 24.5$, the calculated total number is not
strongly dependent on the chosen value for the upper limit of the
integration. Although there will be some FRIIs included, the
majority of sources will be FRIs, therefore integrating out to $\infty$ 
will derive a maximal FRI space density which ought also to be close
to the true value. This calculation gives 460 FRIs/Gpc$^{3}$ brighter 
than $10^{25}$ W/Hz, which is in good agreement with the values calculated by the two direct
measurements.   

\subsection{Results of the space density calculation}

\begin{figure}
\centering
\includegraphics[scale=0.35, angle=90]{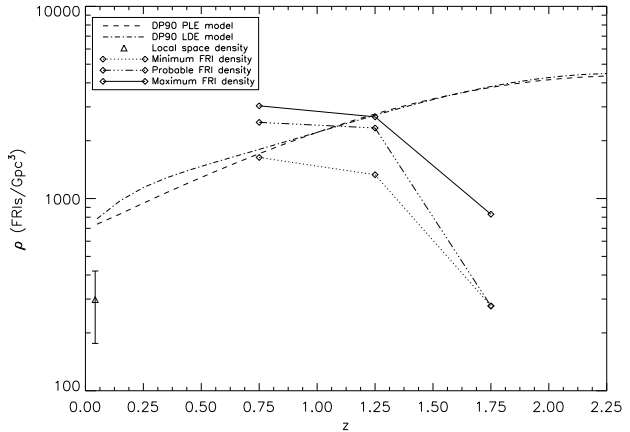}
\caption{\label{dens_fig1} The space density changes in redshift bins
  of width 0.5, plotted at the bin midpoints. Also shown are the PLE and LDE models
  of DP90, converted to 1.4 GHz and to this cosmology, for
  comparison; see \S\ref{dp_bit} for a full description of these models. Since
  no sources with $P>10^{25}$~W/Hz were found in the  
  first redshift bin, no space density value is plotted there.} 
\end{figure}

Figure \ref{dens_fig1} shows the results of the space density
calculation (calculated using Equation \ref{dens_eqn}) for the
4 redshift bins, plotted at the bin midpoints; the numbers are given
in Table \ref{calc_table} and the weights used for each source can be
found in Paper I. 

\begin{table*}
\centering
\caption{\label{calc_table}The results of the space
  density calculations for both redshift bin sizes, for the limiting
  power of 10$^{25}$ W/Hz. `No'. is the number
  of sources in each bin and $\bar{P}$ is the mean luminosity in each
  bin. Also shown are the results of the space density 
  calculations using the additional luminosity limits $P\geq 10^{24}$ and $P \geq
  10^{24.5}$ W/Hz. All results are shown for the maximum, minimum and
  probable FRI numbers.} 
\begin{tabular}{c|c|c|c|c|c|c|c|c|c}
\hline
    & \multicolumn{3}{|c|}{Minimum} & \multicolumn{3}{|c|}{Probable} & \multicolumn{3}{|c}{Maximum} \\
\hline
Bin & No. & $\log{\bar{P}}$ & $\rho$  & No. & $\log{\bar{P}}$ & $\rho$ & No. & $\log{\bar{P}}$ & $\rho$ \\
&&&(FRIs/Gpc$^{3}$) &&&(FRIs/Gpc$^{3}$)&&&(FRIs/Gpc$^{3}$)\\
\hline
\multicolumn{7}{c}{$P \geq 10^{25.0}$ W/Hz} \\
\hline
$0.0 < z \leq 0.5$ & 0 & --   & 0    & 0 & --   & 0    & 0 & --   & 0    \\
$0.5 < z \leq 1.0$ & 3 & 25.7 & 1636 & 4 & 25.6 & 2492 & 5 & 25.5 & 3037 \\
$1.0 < z \leq 1.5$ & 4 & 25.6 & 1331 & 7 & 25.6 & 2329 & 8 & 25.6 & 2662 \\
$1.5 < z \leq 2.0$ & 1 & 26.6 & 276  & 1 & 26.6 & 276  & 3 & 26.3 & 829  \\
\hline                
$0.5 < z \leq 1.5$ & 7 & 25.6 & 1446 & 11 & 25.6 & 2391 & 13 & 25.6 & 2804 \\
\hline
\multicolumn{7}{c}{$P \geq 10^{24.5}$ W/Hz} \\
\hline
$0.0 < z \leq 0.5$ & 1 & 24.7 & 2442 & 2 & 24.6 & 4884 & 3 & 24.7 & 7326 \\
$0.5 < z \leq 1.0$ & 3 & 25.7 & 1636 & 5 & 25.5 & 3370 & 8 & 25.3 & 5006 \\
$1.0 < z \leq 1.5$ & 4 & 25.6 & 1331 & 8 & 25.6 & 2882 & 9 & 25.6 & 3215 \\
\hline
\multicolumn{7}{c}{$P \geq 10^{24.0}$ W/Hz} \\
\hline
$0.0 < z \leq 0.5$ & 1 & 24.7 & 2442 & 3 & 24.6 & 7546 & 4  & 24.6 & 9988 \\
$0.5 < z \leq 1.0$ & 3 & 25.7 & 1636 & 5 & 25.5 & 3370 & 10 & 25.3 & 6892 \\
\hline
\end{tabular}
\end{table*}

The space densities for the minimum, probable and maximum FRI groups
show a high redshift comoving density enhancement for the FRIs in this
sample, compared to the local FRI space density. The turnover in space
density seen at $z \gtrsim 1.5$ is supported by 
the work of \citet{waddington2} who found evidence for a high redshift
cut--off for their lower luminosity radio sources by $z \simeq 1$ --
$1.5$. However, since the values of $V_{i}$ in the last bin can depend
on assumed spectral indices, the addition of a small number of steep
spectrum sources, close to the luminosity limit, would result in a large increase in density;
for example, \squig\ 3 more sources of flux density $S_{\rm lim}$ and
$\alpha = 1.0$ would be needed to give the same maximum value of
$\rho$ as in the previous $1.0 < z < 1.5$ bin. Sources may also have
been missed in this bin because of the strong observational bias
against FRIs at these distances (i.e. a clear Group 1 source at $z =
1$ could be classified as a Group 3 source if it was at $z=2$). 

\begin{figure}
\centering
\includegraphics[scale=0.35, angle=90]{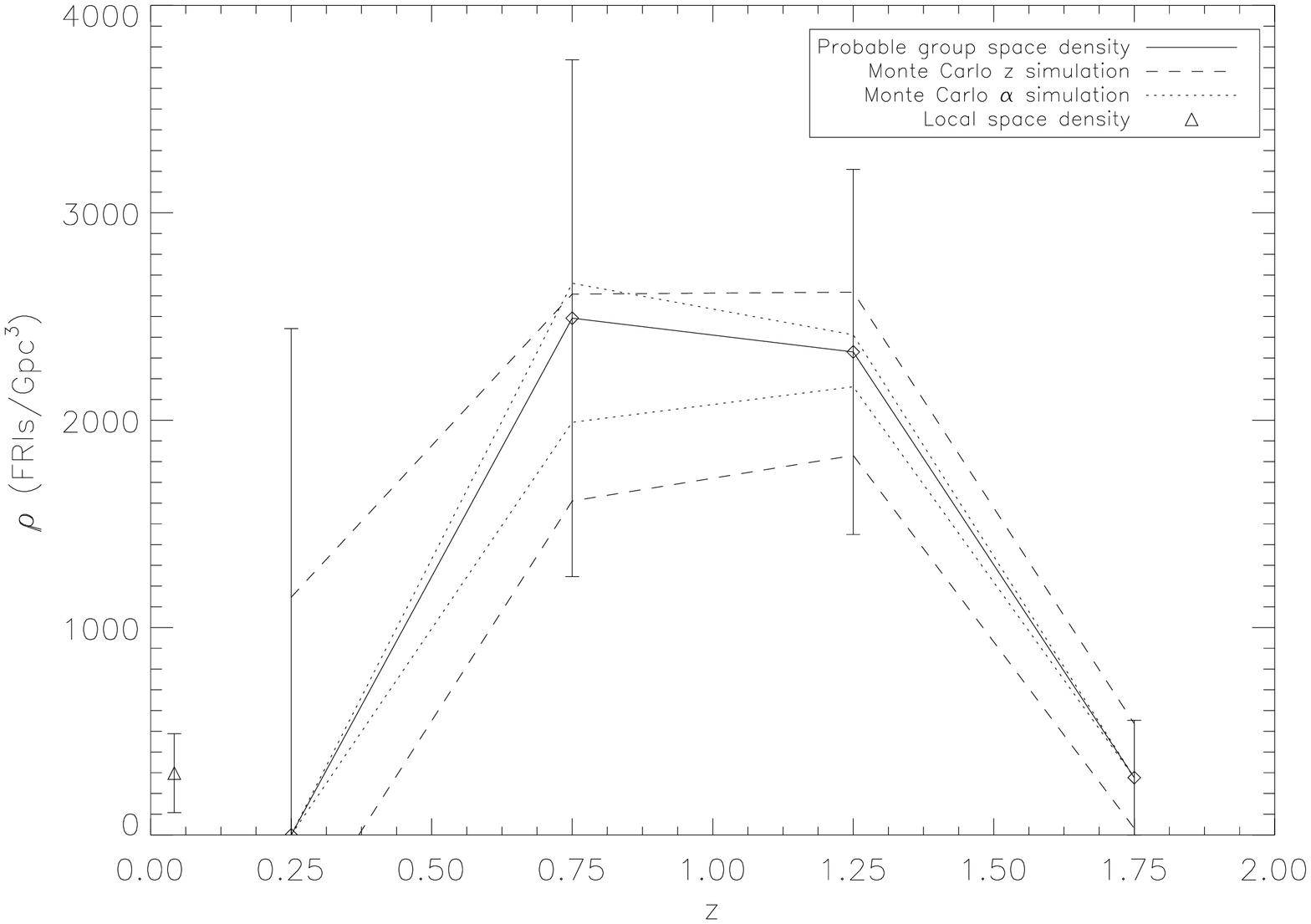}
\caption{\protect\label{mc_fig} The space density changes and Poisson errors of the probable
  group of objects, with $P \geq 10^{25}$ W/Hz, overplotted with the $\pm$1$\sigma$ results of the
  redshift and spectral index Monte Carlo simulations. Both sets of results clearly
  show that it is the small number of sources in each bin that has the most effect on the results.} 
\end{figure}

Two potential concerns in calculating the space densities are the
estimated redshifts\footnote{A full description of the redshift
  estimation methods used, and the results obtained for those sample
  sources without a secure redshift, can be found in Paper I} and the assumed
spectral indices of 0.8 used for some of the sample sources. Error in 
both of these factors could affect the radio power determination, and lead to sources
falsely falling below $P_{\rm lim}$, whilst error in the former alone
could also move sources between redshift bins. To 
investigate the effect of these on the results, two Monte Carlo 
simulations, with 10,000 iterations each, were performed. In the first simulation, in each iteration,
the redshifts which were estimated were varied by a factor drawn randomly from a Gaussian
distribution of width equal to 0.2 in $\log z$, the approximate spread
in both the r--z and K--z relations. Similarly, in the second
simulation, in each iteration, the assumed spectral indices were
varied from 0.8 by a factor drawn randomly from a Gaussian
distribution of width 0.5. This value was chosen as it represents a
reasonable spread in $\alpha$. Figure \ref{mc_fig} shows the
$\pm$1$\sigma$ density results of the two simulations, carried out for the probable FRIs 
(i.e. groups 1 and 2) only. Also shown are the errors on the densities
calculated from the simple Poisson errors on the number of sources in
each bin (for the bin containing no sources, an error of $\pm$1 source
was assumed). These results are also given in numerical form in Table
\ref{mc_table}. It is clear from this that the major limiting factor for the results is the small 
number of sources in the sample, rather than the redshift estimates,
or the assumed spectral indices, that were used for some of the sources. It should also be noted that
the radio FRI classifications are as comparably large a source of error
as the redshift estimates.  

\begin{table*}
\centering
\caption{\label{mc_table} The errors calculated from the Monte Carlo
  simulations, $z ~ \sigma_{\rm mc}$ and $\alpha ~ \sigma_{\rm mc}$, along with the Poissonian,
  $\sigma_{\rm p}$, errors and the spread in the calculated space
  densities for each redshift bin ($\rho_{\rm max} - \rho_{\rm min}$); all values are in
  FRIs/Gpc$^{3}$. Also shown is the probable FRI space density,
  $\rho_{\rm prob}$.}
\begin{tabular}{c|c|c|c|c}
\hline
                  & $0.0 < z \leq 0.5$ & $0.5 < z \leq 1.0$ & $1.0 < z
\leq 1.5$ & $1.5 < z \leq 2.0$ \\ 
\hline
$\rho_{\rm prob}$                    & 0    & 2492 & 2329 & 276  \\ 
\hline
$\rho_{\rm max} - \rho_{\rm min}$    & 0    & 1401 & 1331 & 553  \\
$z  ~ \sigma_{\rm mc}$                  & 810  & 496  & 391  & 257  \\
$\alpha ~ \sigma_{\rm mc}$             & 0    & 345  & 124  & 0    \\
$\sigma_{\rm p}$                     & 2442 & 1246 & 880  & 276  \\
\hline
\end{tabular}
\end{table*}

\subsubsection{Quantifying the space density enhancement \label{dp_bit}}

The calculations above show that the uncertainty in the space
density results is dominated by the Poissonian error. In order to
better quantify the high redshift enhancements therefore, the density
calculation was repeated using a single, large, redshift bin spanning
$0.5 < z < 1.5$, containing a minimum of 7 and a maximum of 13 FRI sources. The resulting
densities can also be found in Table \ref{calc_table}; they show
enhancements of 3.2~$\sigma$, 2.9~$\sigma$ and 2.1~$\sigma$ over the
local FRI space density, for the maximum, probable and minimum groupings respectively.

Measuring these significance levels is not the best way of judging the
reliability of the space density increase, however. A better method,
following \citet{Ignas}, is to assume no evolution of 
the FRI population and then calculate the probability of detecting the
numbers of minimum ($P_{>7}$), probable ($P_{>11}$) and maximum ($P_{>13}$)
objects seen in the $0.5 < z < 1.5$ bin, if this no--evolution scenario was
correct. The volume contained within this bin is 0.005~Gpc$^{3}$ which,
assuming no evolution occurs from a local density of 298
FRIs/Gpc$^{3}$, \ gives an expected number of 1.44 FRIs over this
range. (For comparison, repeating this calculation for $z < 0.5$
suggests that 0.12 FRIs should have been detected in the sample, in this volume, for a constant co--moving 
space density.) The resulting probabilities are summarized in
Table \ref{prob_table}. These are all $\ll$1\% which indicates that, as
expected, the no--evolutionary scenario can be discounted. 

\begin{table}
\centering
\caption{\label{prob_table} The probabilities calculated for the
  no--evolution, PLE and LDE scenarios for the minimum, probable and
  maximum numbers of FRIs.} 
\begin{tabular}{c|c|c|c|c}
\hline
             &Expected & $P_{>7}$ & $P_{>11}$ & $P_{>13}$ \\
& Number & & & \\
\hline
No Evolution &  1.44          & 0.07\%   & 3.8$\times$10$^{-5}$\% & 4.9$\times$10$^{-7}$\% \\
PLE          &  11.63         & 94\%     & 61\%                   & 38\% \\ 
LDE          &  11.67         & 95\%     & 62\%                   & 39\% \\
\hline
\end{tabular}
\end{table}

The pure luminosity evolution (PLE) and luminosity/density
evolution (LDE) models of DP90, which fit the overall radio source population well out
to $z \sim 2$, can also be compared to the results using this 
method. If the behaviour of the FRIs here is consistent with these PLE
and LDE models, this would suggest that they evolve in the same way as  
FRII objects of the same radio power. 

In the PLE model, the local radio luminosity function (fitted using
a dual power--law, with a steeper slope, for the higher powered
sources, above the luminosity break) shifts horizontally in the
$\rho$--$z$ plane only, and its overall shape does not change. These
redshift changes were confined to the luminosity normalization,
$P_{\rm c}$, only and DP90 parameterized these as a quadratic in $\log
z$. Conversely, in the LDE model, the local RLF can move both
horizontally and vertically and, therefore, can steepen or flatten. As a result,
DP90 allowed the density normalization, $\rho_{0}$, to also vary with
redshift. For full details of these two models, the reader is referred
to Section 3.4 of DP90. 

For each of these two models, the total number of sources in
the interval $0.5 < z < 1.5$ was found by integrating  
\begin{eqnarray}
N_{\rm tot} &=& \Omega \int^{1.5}_{0.5} \int^{P2}_{P1} \rho(P, z)~dV(z)
\label{dp_int}
\end{eqnarray}
where $\Omega$ is the area of the survey in steradians and the
luminosity limits of the integration are $P1 = 10^{25}$ W/Hz and $P2 =
10^{27.5}$ W/Hz. The co--efficients used in these calculations were
those determined for the steep spectrum population by DP90. Where necessary these values
were converted from their original 2.7 GHz to 1.4 GHz, the frequency
used here, again assuming an average $\alpha$ of 0.8. Changing 
this assumption does not significantly alter the final
results. Additionally, since the DP90 models were determined using
Einstein de Sitter (EdeS) cosmology, the luminosity limits actually used for
the integration were the EdeS values corresponding to $P_{1}$ and
$P_{2}$ at each redshift step to ensure that the correct, concordance
cosmology, values of $N_{\rm tot}$ were found. 

Carrying out the integrations results in $N_{\rm tot}
\sim 11$ for both models, over this redshift range, and corresponding
enhancements over the local value used here of \squig\ 8. This
expected number agrees very well with the observed numbers of 7, 11
and 13 for the minimum, probable and maximum samples, confirming that these
space densities can be consistent with these models. The subsequent
probabilities arising from these numbers can again be found in Table
\ref{prob_table}. This result is further supported by Figure \ref{dens_fig1}, in which the density values
predicted by the PLE and LDE models are overplotted with the results
previously calculated for the 4 redshift bins. The low redshift disagreement seen between the
models and the local FRI space density is likely to arise from the poor constraints
locally, at these radio powers, in the DP90 results. 

\subsubsection{Including the lower luminosity FRIs}

\begin{figure}
\centering
\includegraphics[scale=0.35, angle=90]{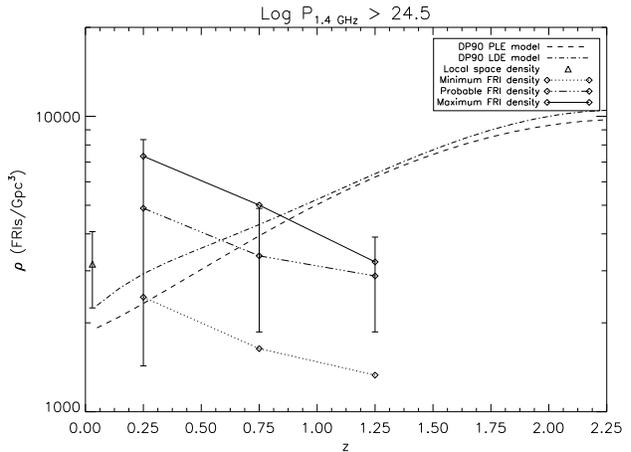}
\caption{\label{245_dens} The space density changes in redshift bins
  of width 0.5, plotted at the bin midpoints for P$>10^{24.5}$  W/Hz. For
  comparison the redshift axis is plotted with the same range as in
  Figure \ref{dens_fig1} and the PLE and LDE models of DP90
  are again shown, along with Poisson errors for the probable group of
  objects.}
\end{figure}

\begin{figure}
\centering
\includegraphics[scale=0.35, angle=90]{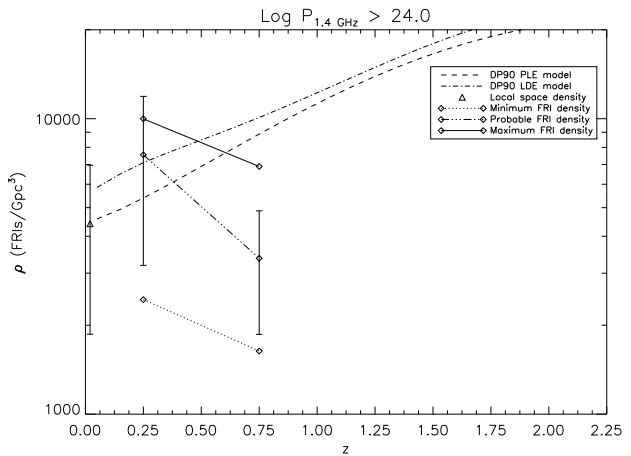}
\caption{\label{24_dens} The space density changes in redshift bins
  of width 0.5, plotted at the bin midpoints for P$>10^{24}$ W/Hz. For
  comparison the redshift axis is plotted with the same range as in
  Figure \ref{dens_fig1} and the PLE and LDE models of DP90
  are again shown, along with Poisson errors for the probable group of
  objects.}
\end{figure}

The space density calculation can be repeated using different
luminosity limits to investigate the behaviour of the weaker sources
in the sample. Figures \ref{245_dens} and \ref{24_dens} show the
results of this for limits of 10$^{24}$ and 10$^{24.5}$ W/Hz and the
density values are given in Table \ref{calc_table}. The calculation
was, in both cases, carried out for the minimum, maximum and probable
numbers of FRI, using redshift bins of width 0.5, with the final bin
dependent on the value of the limiting luminosity as discussed
previously. The local values were again found using a combination of
the 3CR and equatorial galaxy samples.   

The comoving space densities for these two new limits indicate that the
enhancement over the local values in both cases is smaller than that seen when $P_{\rm lim} = 10^{25}$ W/Hz (Figure
\ref{dens_fig1}). These results are supported by the work of Sadler et
al. \citeyearpar{Sadler} who found that the cosmic evolution of their
low power ($10^{24} \leq P_{\rm 1.4 GHz} < 10^{25}$ W/Hz) population
was significant but less rapid than that seen for their higher powered
sources. 

Inspection of Figures \ref{245_dens} and \ref{24_dens} also suggests that the peak
comoving density moves to lower redshifts at lower luminosities, though the
large errors on these results, and the restricted redshift range
sampled at these low powers prohibit strong conclusions from being
drawn.  

\section{Summary and discussion}
\label{discuss}

The radio observations presented here, along with the data
previously presented in Paper I, allowed the 81 sources of the
complete sample to be classified into 5 groups: Group 1 was for
secure, morphologically classified, FRIs; Group 2 for sources with indications
of FRI--type extension, alongside a flux density loss of 3$\sigma$ or
more; Group 3 for sources with either flux density loss or some
extension, and Groups 4 and 5 for sources which were either definitely
not FRIs or which were unclassifiable. The FRIs in the sample, together with the two FRIs in the Hubble Deep and
Flanking Fields, were then used to calculate the maximum, probable and
minimum space densities over the range $0.0 < z < 2.0$. The main
results of this calculation are listed below.
\begin{enumerate}
\item Clear density enhancements, by a factor 5--9, were seen at $z \sim 1.0$, over the
local FRI value, for FRIs brighter than $10^{25}$ W/Hz,
\item This result is secure, with the main source of error arising
  from the low number of sources in the FRI subsample,
\item The no--evolution scenario was ruled out at the $>$99.9\% confidence limit,
\item The results were consistent with previously published models
  which in turn suggests that, at a particular radio power, FRIs evolve like FRIIs.
\item The results also indicate that the FRI evolution may be luminosity
  dependent, with lower powered sources evolving less strongly, and
  the peak in space density possibly moving to lower
  redshifts for weaker objects. 
\end{enumerate}

Now that the space density enhancements of this sample have been
determined it is useful to compare them to the results of previous
work. Sadler et al. \citeyearpar{Sadler}, for instance, found
increases in space density, out to $z = 0.7$, by a factor of \squig\ 2
-- 10 for the high ($P_{1.4 GHz} > 10^{25}$ W/Hz) luminosity radio galaxies in their sample; these
are consistent with the range of enhancements (\squig\ 5 -- \squig\ 9)
to $z \sim 1$ seen in the large redshift bin here. Jamrozy et al. \citeyearpar{Jamrozy}
also find that positive evolution, this time PLE, of the form $\rho(z)
= \rho(0) \exp (M(L)\tau)$ (where $\tau = 1 - (1+z)^{-1.5}$ and the
evolution rate, $M(L)$, was found to be 5.0), is necessary to fit the
number counts of the highest luminosity, morphologically selected, FRIs in their sample. At $z=1$ 
this corresponds to an enhancement of 25 which is significantly higher
than that seen here. Additionally, Willott et al. \citeyearpar{Willott} see
a rise of about one dex, between $z \sim 0$ and $z \sim 1$, in the comoving space density of their low
luminosity, weak emission line, population which contains mainly FRI sources, along with
some FRIIs. It should be noted that other studies of radio galaxy
evolution do not directly measure the amount of space density evolution
present in their samples. Instead they test for evolution using the
$V/V_{\rm max}$ statistic and therefore cannot be directly compared to the
results of this work. 

In general, all the previous studies of radio galaxy evolution
conclude that the more luminous sources undergo more cosmic
evolution. For samples in which the FRI population was determined by a
luminosity cut, this tends to imply little or no evolution for these
sources (e.g. the lower luminosity sources of Clewley \& Jarvis (2004)
show no evolution, whereas evolution is seen for sources of comparable
luminosity to those here). However, for samples which either morphologically classify
their sources or which apply a different dual--population scheme
(e.g. the low/high luminosity division based on line luminosity of Willott et
al. \citeyearpar{Willott}), evolution of at least some of the FRIs objects is typically seen. It
would seem, therefore, that the previous models in which all FRIs have
constant space density, whereas all FRIIs undergo strong, positive,
cosmic evolution, are too simplistic and do not accurately represent
the behaviour of the FRI--type objects. A better representation of the
FRI/II space density evolution seen here and in other work, is the
picture in which as the luminosity of a source increases, so does the
amount of positive evolution it undergoes between z \squig\ 0 and
1--2. It should be noted though that the behaviour of the low power
FRIs is not throughly investigated here, but the evolution seen for higher
power objects is consistent with this model. 

The detection in this work of FRI space density enhancements for sources with
$P_{\rm 1.4GHZ} \geq 10^{25}$ W/Hz, along with the previous studies which
find essentially luminosity dependent evolution for the FRI
population, strongly suggests that neither the intrinsic or the
extrinsic difference models can fully explain the observed FRI/II
differences. When radio galaxies are divided according to their line
luminosities, as in Willott et al. \citeyearpar{Willott}, there are both FRI
and FRII sources in the low excitation, low luminosity, population;
this implies that these FRIIs may also have inefficient accretion
flows similar to those previously proposed for FRIs (e.g. Ghisellini
\& Celotti 2001). Jets produced by low accretion flow sources are
generally weak with the majority having an FRI type structure, whereas
higher accretion flows give rise to stronger, mainly FRII type
jets. In this scenario therefore, both low and high accretion flows 
are capable of producing both FR jet structures and the morphological
differences between the two classes can be explained extrinsically, as
a function of their individual environments, whilst the differences in
line luminosity can be explained intrinsically, as a function of their
black hole properties \citep[e.g.][]{hardcastle2}.   

In summary, therefore, combining these results with the previously published work in this
field leads to the conclusion that the intrinsic/extrinsic models used
to describe the differences in radio sources are too simplistic, and that what
is needed is some combination of the two. Dividing radio galaxies
according to line luminosity, instead of according to their observed
morphology, may be the answer as, in this classification, weak,
inefficiently accreting, FRIIs can be grouped together with FRIs.  

\section*{Acknowledgments}

EER acknowledges a research studentship from the UK Particle Physics
and Astronomy Research Council. PNB would like to thank the Royal Society for generous financial support
through its University Research Fellowship scheme. The VLA is operated
by the National Radio Astronomy Observatory, a facility of
the National Science Foundation operated under cooperative agreement
by Associated Universities Inc. MERLIN is a National Facility
operated by the University of Manchester at Jodrell Bank Observatory
on behalf of STFC   

{}

\bsp

\appendix

\section{Lynx B--array radio images}

\begin{figure*}
\hspace{-25mm}
\begin{minipage}{15cm}
\includegraphics[scale=0.85]{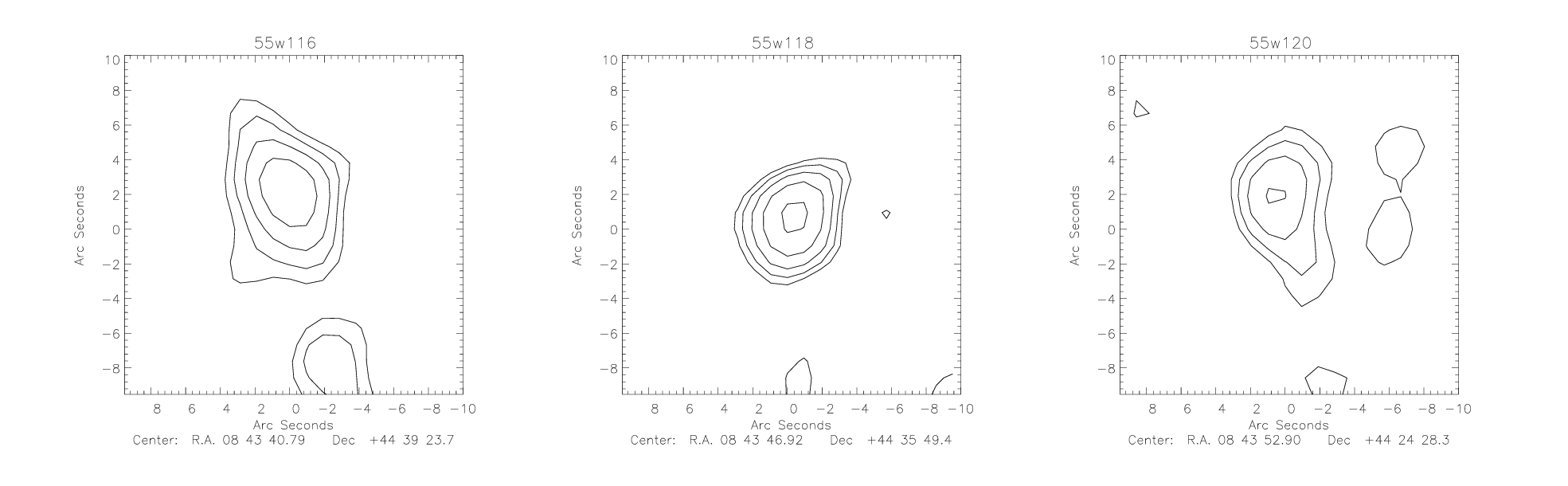}\vspace{-1mm}
\includegraphics[scale=0.85]{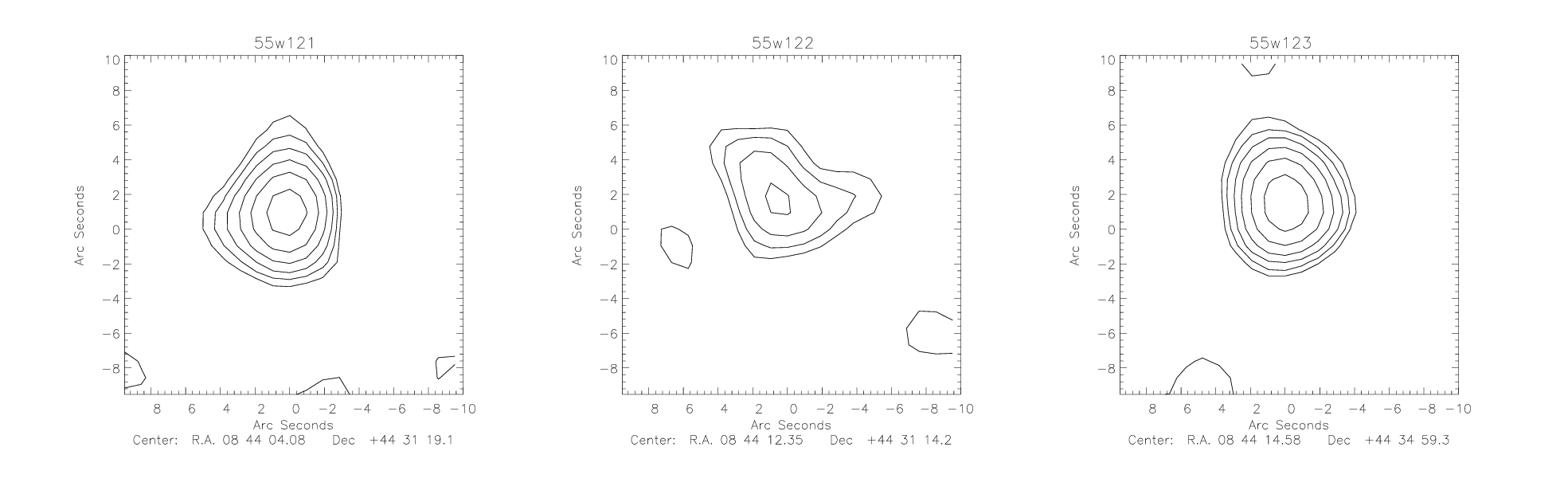}\vspace{-1mm}
\includegraphics[scale=0.85]{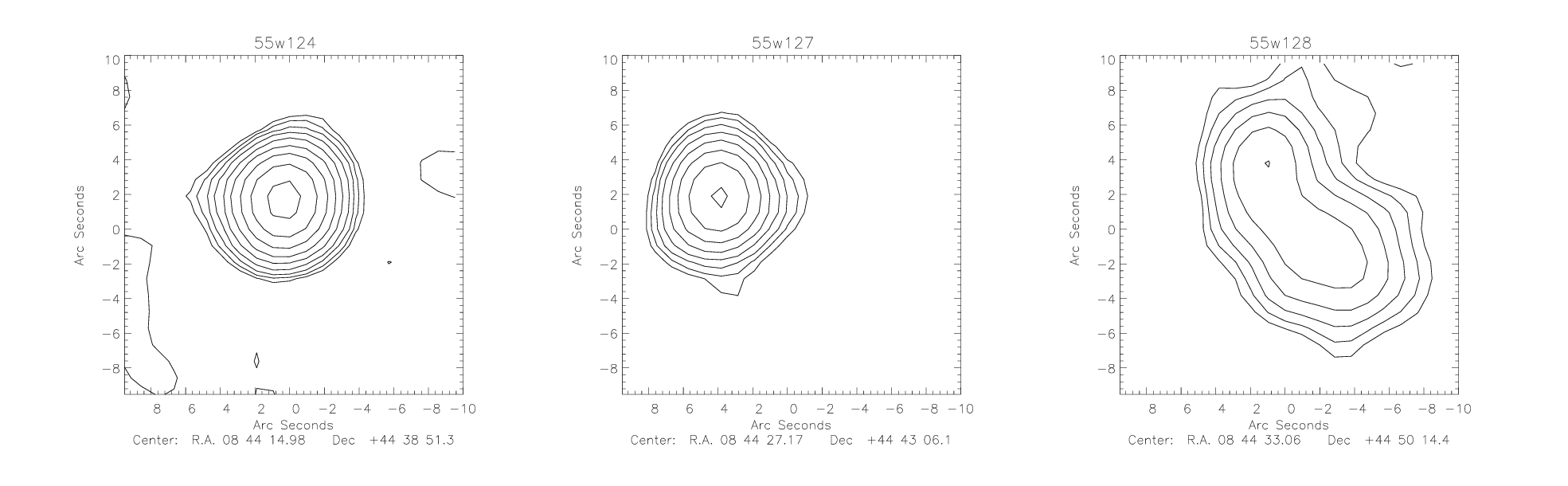}\vspace{-1mm}
\includegraphics[scale=0.85]{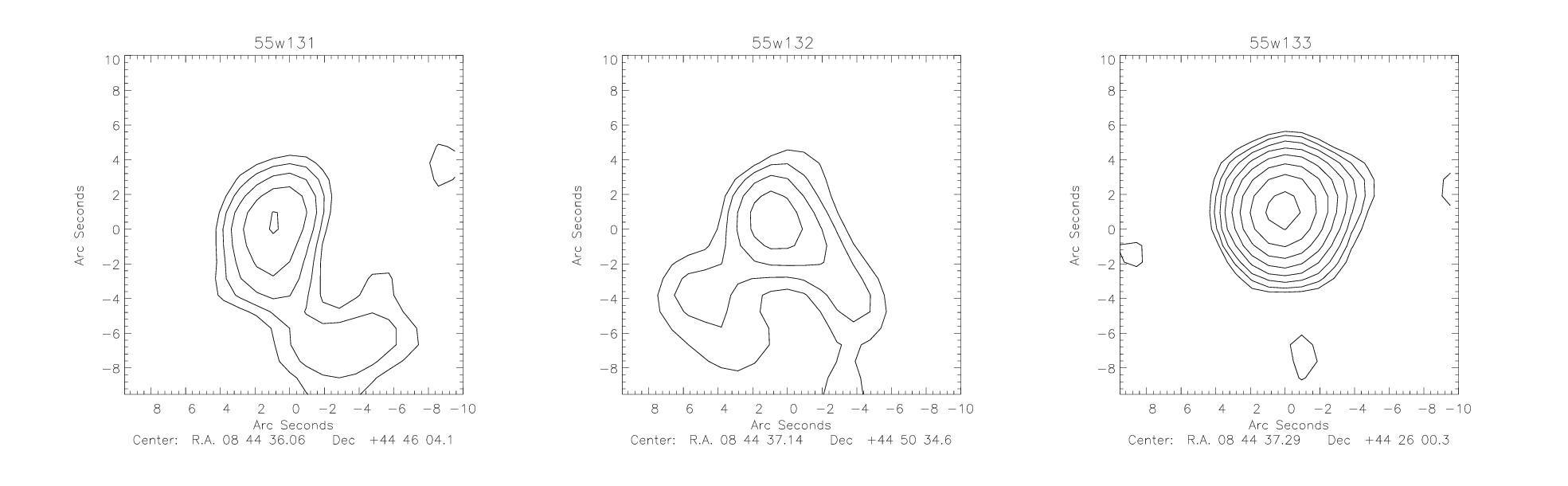}\vspace{-1mm}
\vspace{5mm}
\end{minipage}
\caption{The radio contour images, for the
  Lynx field complete sample, from the VLA 1.4GHz B--array
  observations. The beam size is 
  5.36\arcsec$\times$4.67\arcsec. Contours start at 50$\umu$Jy/beam and are
  separated by factors of $\sqrt{2}$. The images are centred on the
  optical host galaxy positions from Paper I if
  available. \protect\label{b_images}} 
\end{figure*}

\begin{figure*}
\hspace{-25mm}
\begin{minipage}{15cm}
\includegraphics[scale=0.85]{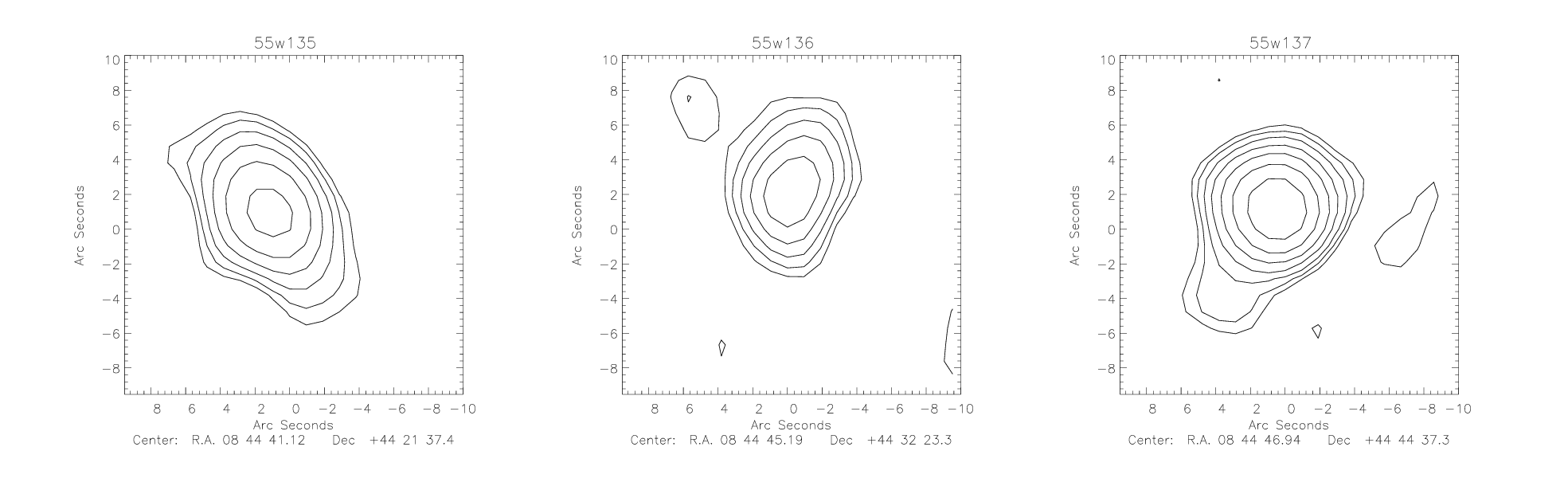}\vspace{-1mm}
\includegraphics[scale=0.85]{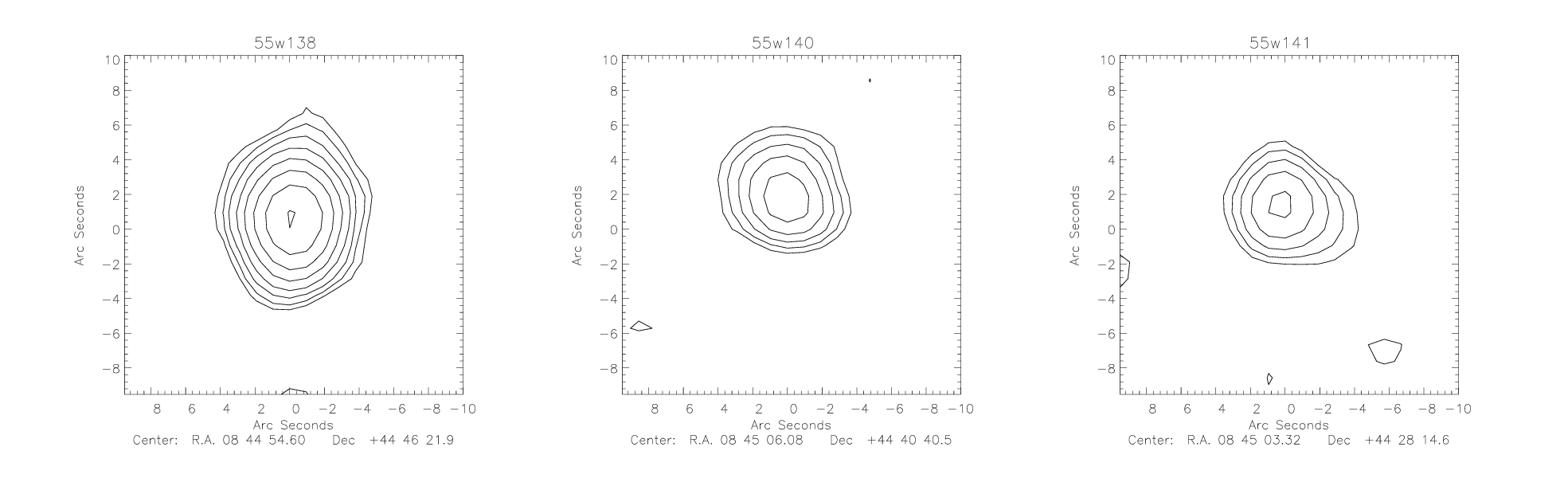}\vspace{-1mm}
\includegraphics[scale=0.85]{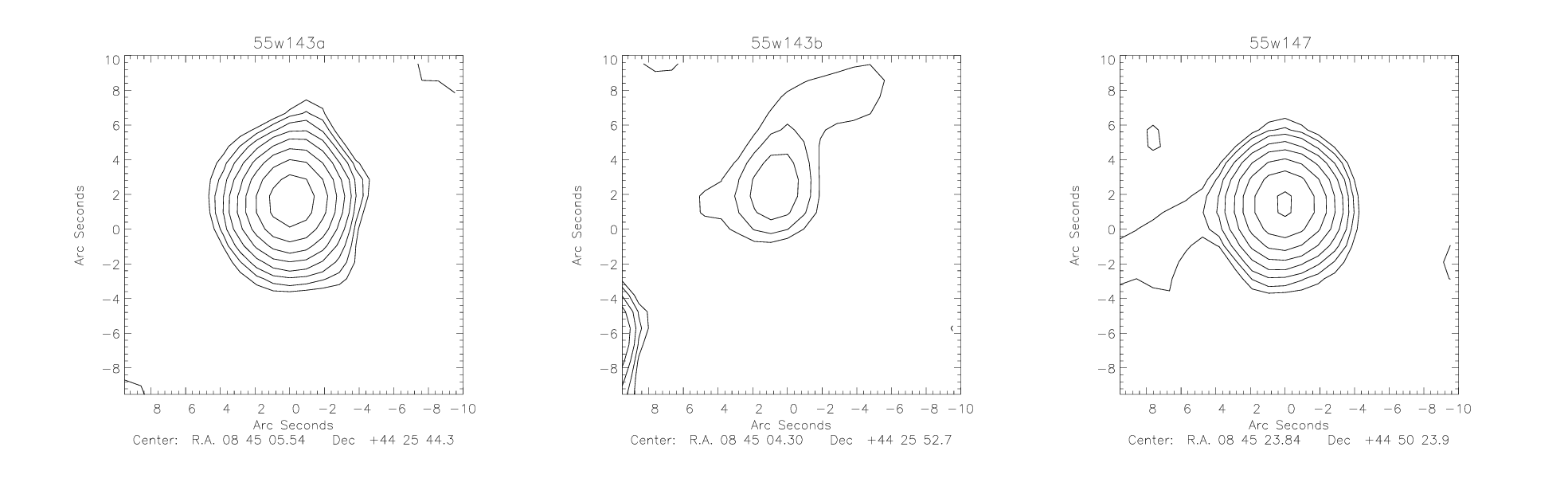}\vspace{-1mm}
\includegraphics[scale=0.85]{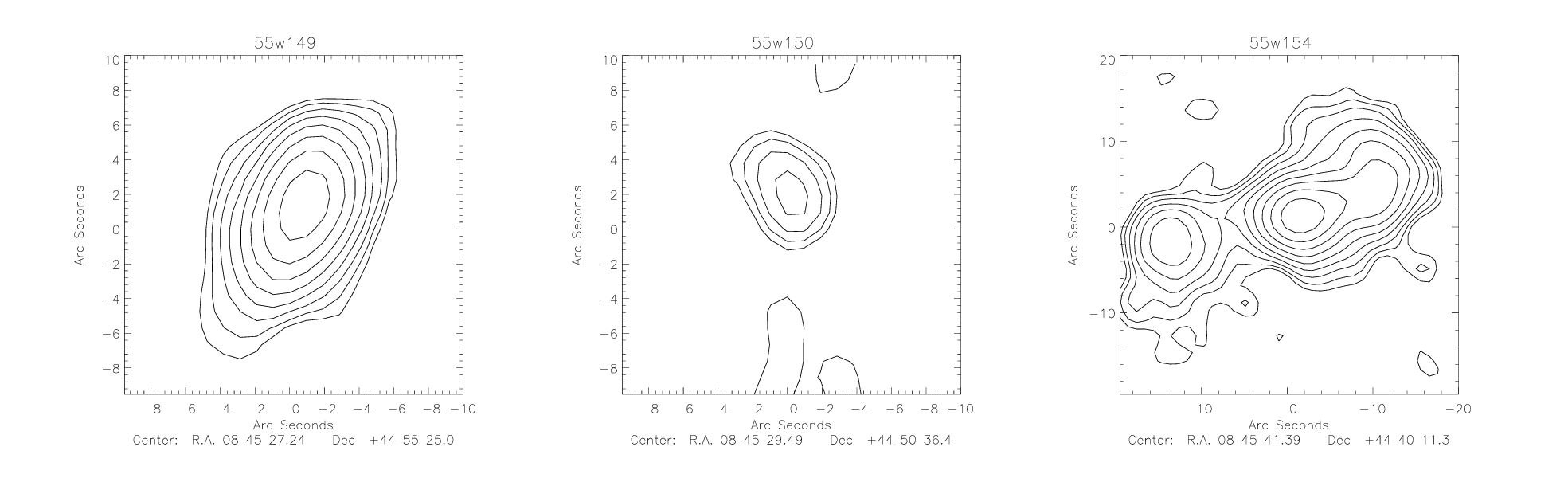}\vspace{-1mm}
\vspace{5mm}
\contcaption{}
\end{minipage}
\end{figure*}

\begin{figure*}
\hspace{-25mm}
\begin{minipage}{15cm}
\includegraphics[scale=0.85]{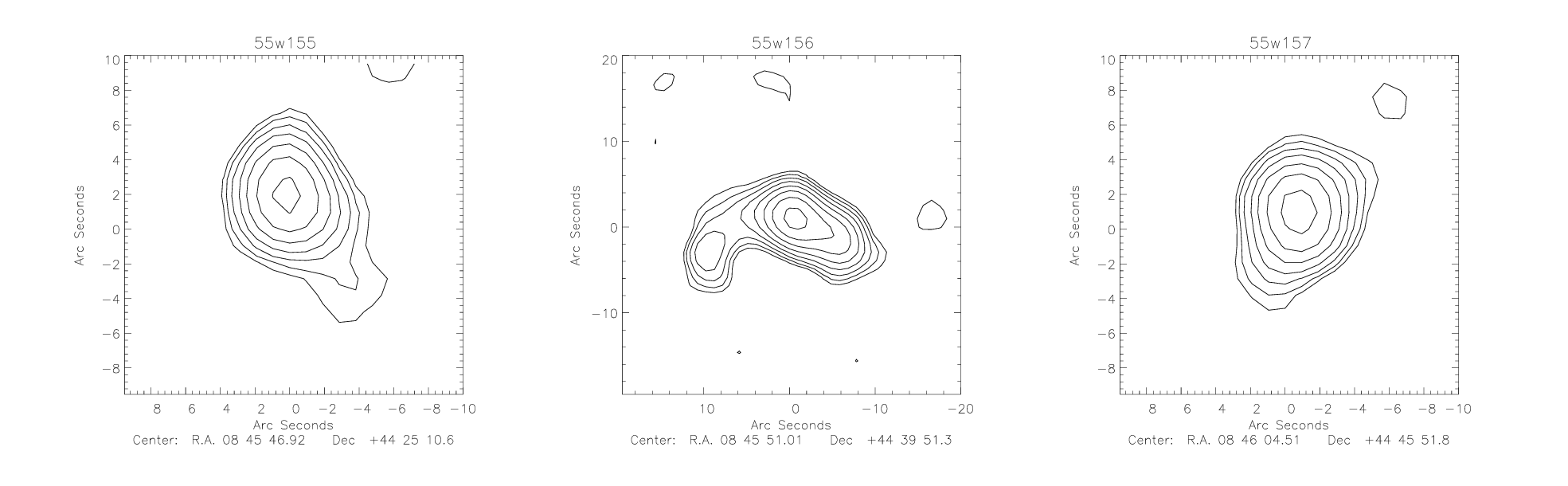}\vspace{-1mm}
\includegraphics[scale=0.85]{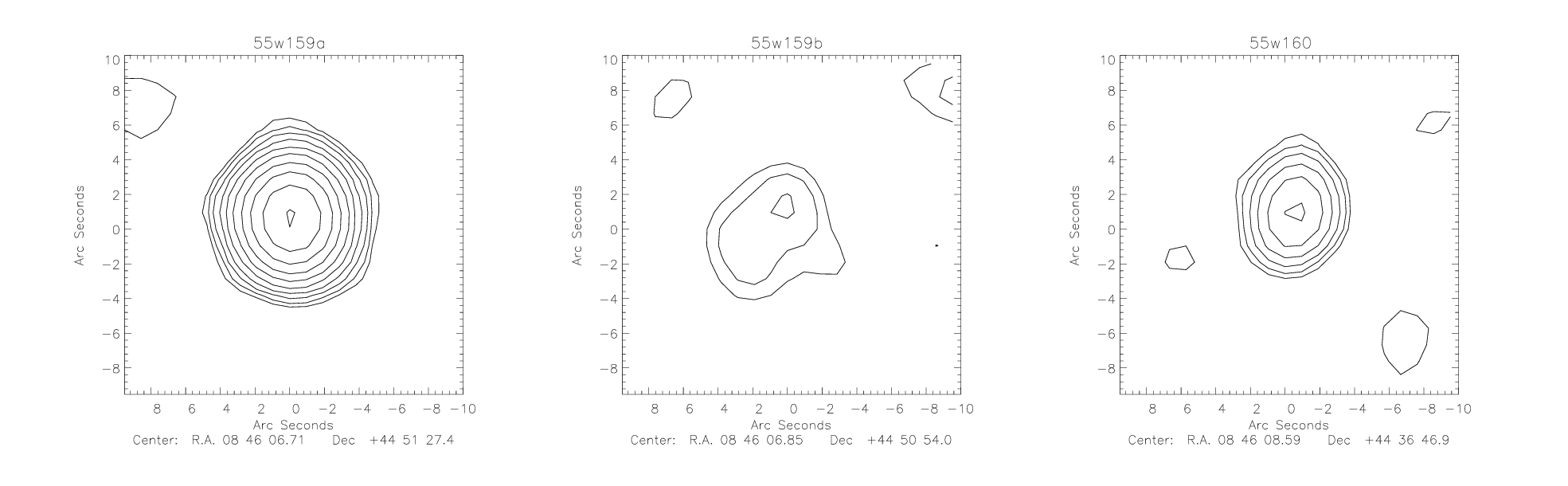}\vspace{-1mm}
\includegraphics[scale=0.85]{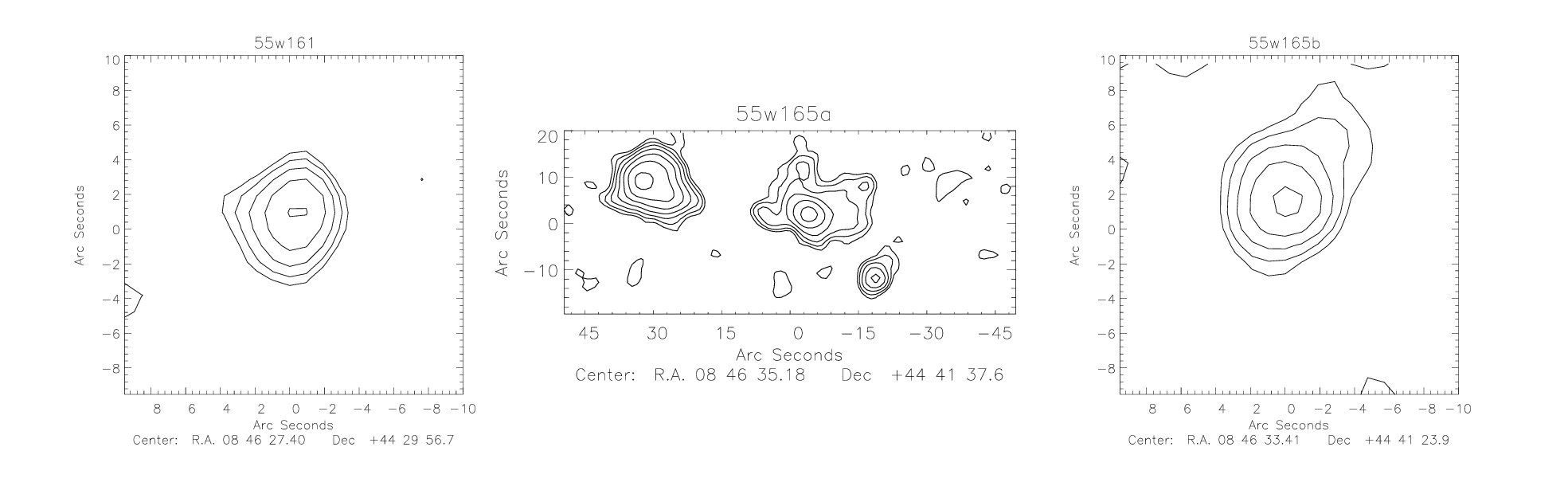}\vspace{-1mm}
\includegraphics[scale=0.85]{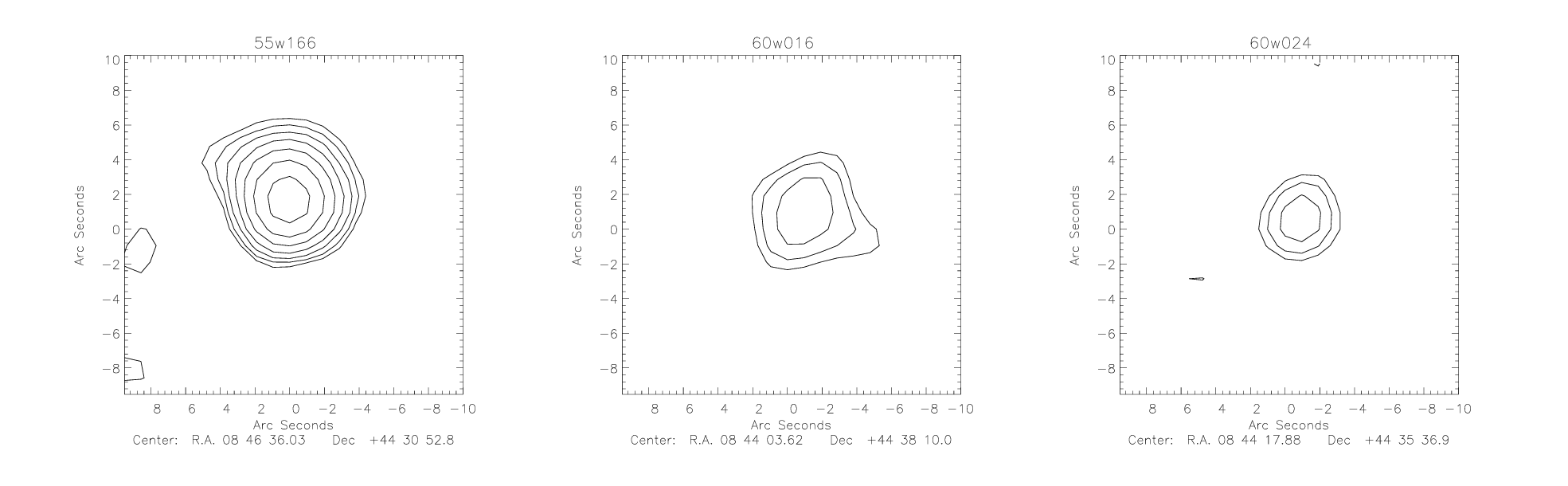}\vspace{-1mm}
\vspace{5mm}
\contcaption{}
\end{minipage}
\end{figure*}

\begin{figure*}
\hspace{-25mm}
\begin{minipage}{15cm}
\includegraphics[scale=0.85]{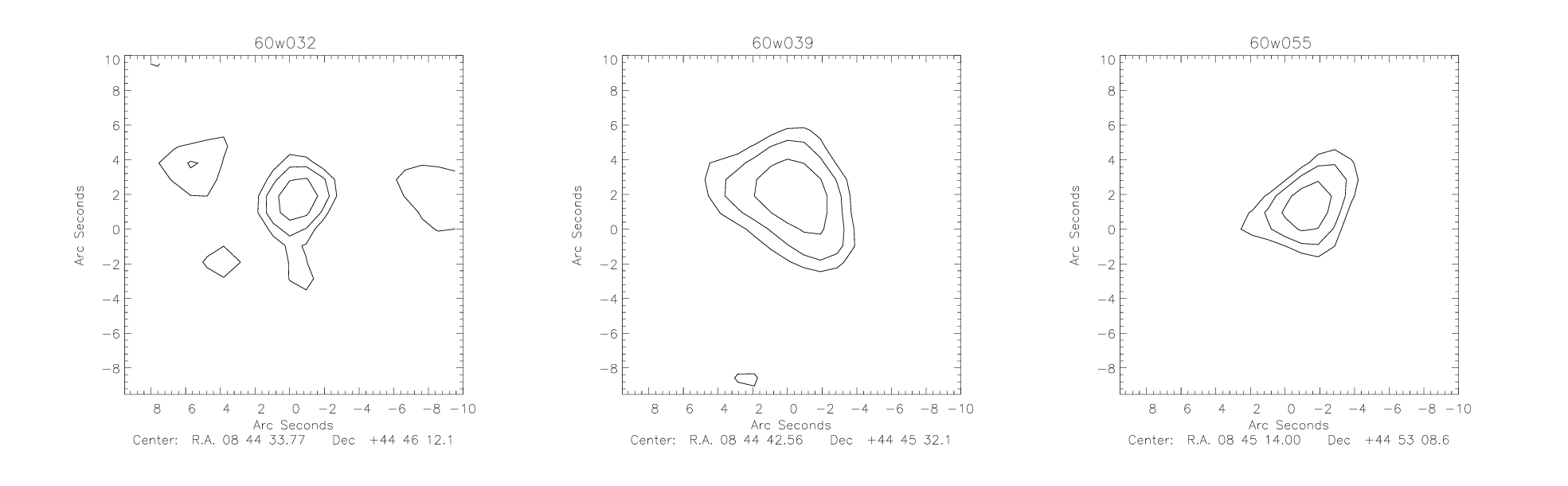}\vspace{-1mm}
\includegraphics[scale=0.85]{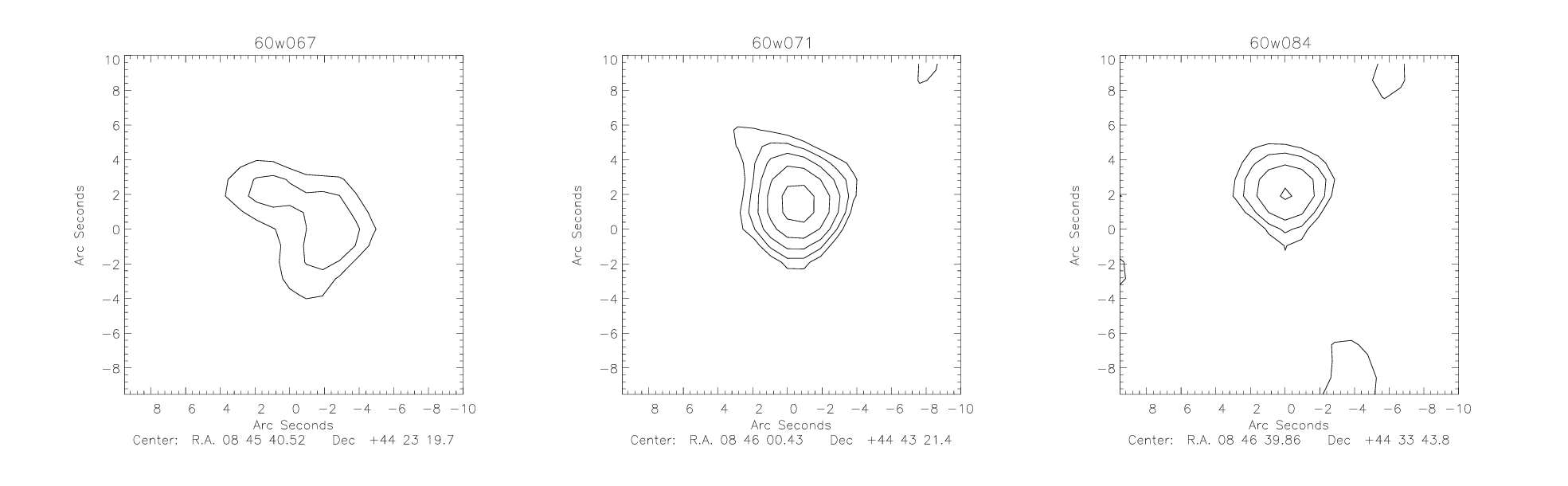}\vspace{-1mm}
\vspace{5mm}
\contcaption{}
\end{minipage}
\end{figure*}

\begin{figure*}
\hspace{-25mm}
\begin{minipage}{15cm}
\includegraphics[scale=0.85]{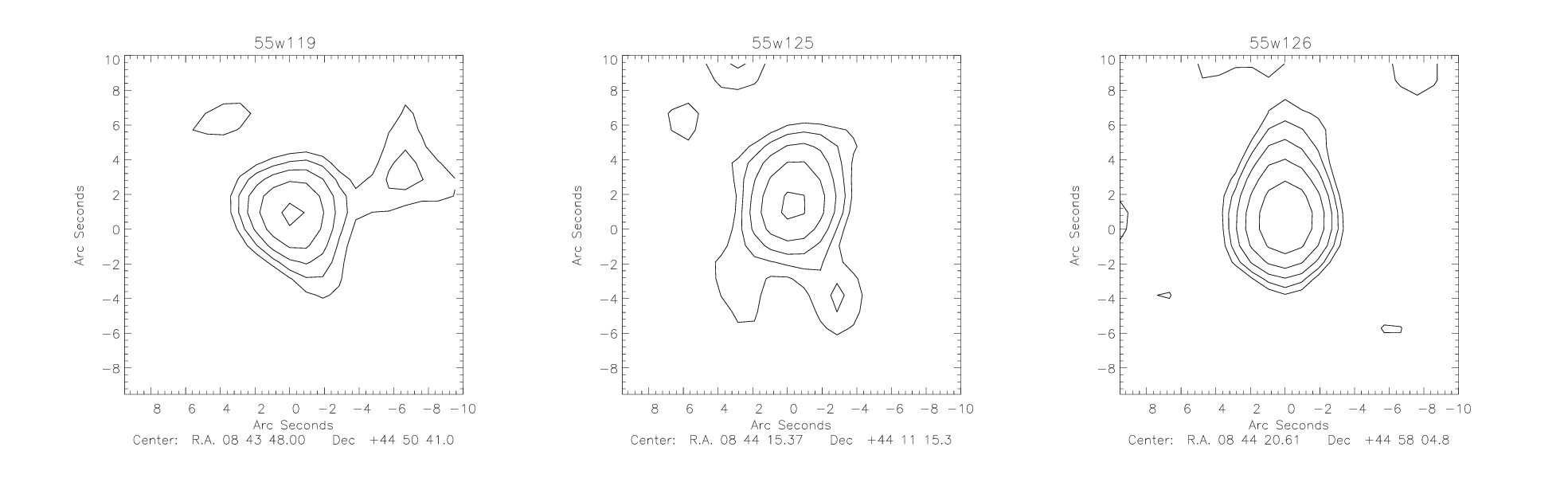}\vspace{-1mm}
\vspace{5mm}
\end{minipage}
\caption{The B--array radio contour images, for
  the sources not included in the 
  Lynx field complete sample. The beam size is
  5.4\arcsec$\times$4.5\arcsec. Contours start at 50$\umu$Jy/beam and are
  separated by factors of $\sqrt{2}$. The images are centred on the
  optical host galaxy positions from Paper I if
  available. \protect\label{rad_b_im_no}} 
\end{figure*}

\section{Complete radio images for sources included in the high--resolution observations}

\begin{figure*}
\hspace{-15mm}
\begin{minipage}{15cm}
\includegraphics[scale=0.85, angle=0]{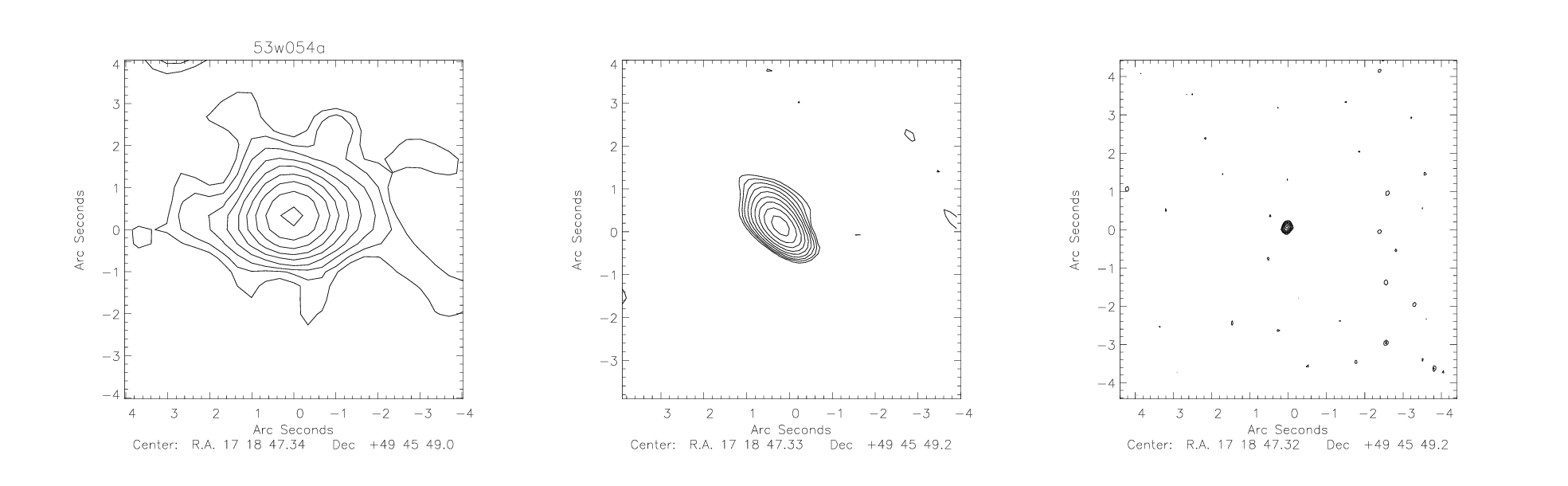}
\includegraphics[scale=0.85, angle=0]{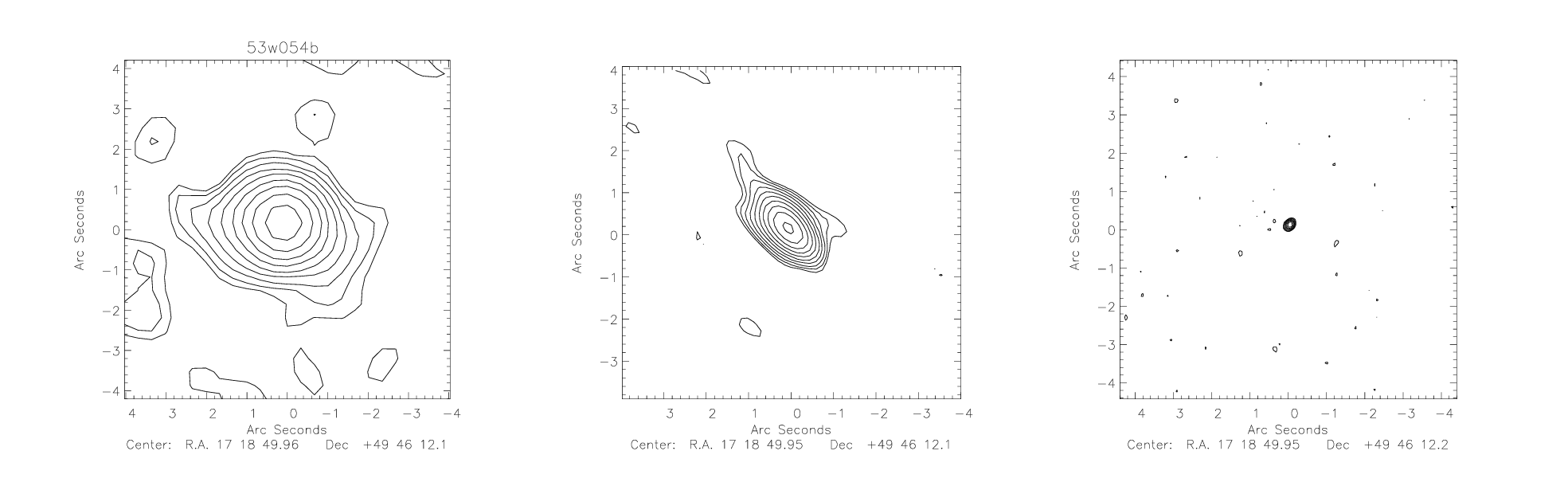}
\includegraphics[scale=0.85, angle=0]{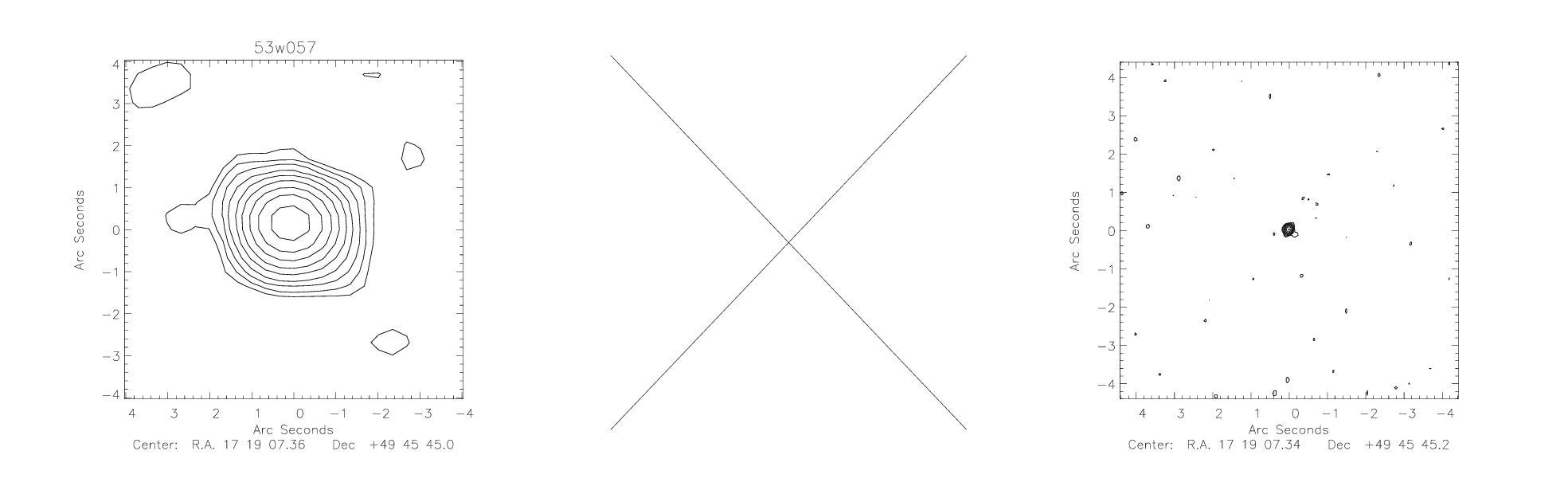}
\vspace{10mm}
\includegraphics[scale=0.85, angle=0]{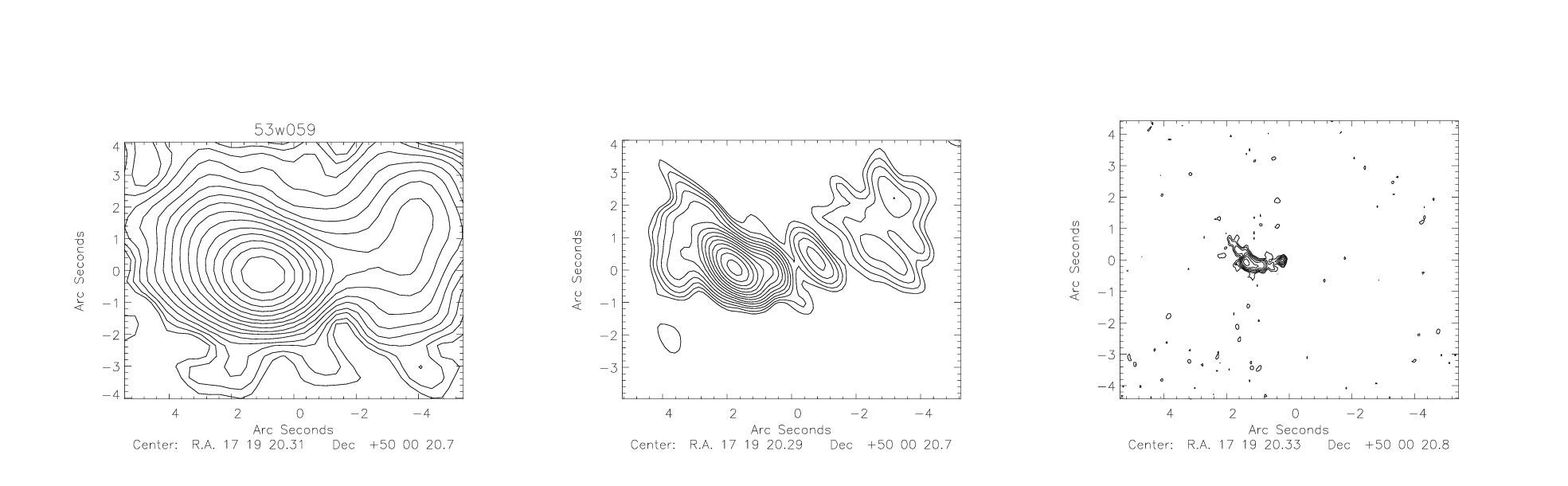}
\end{minipage}
\caption{The A--array, A+Pt and MERLIN contour maps for the sources in 
the Hercules field, included in the A+Pt or MERLIN observations, all
centred on the optical host galaxy position if available, with the
aim of comparing the source morphologies at the three different
resolutions, $\sim$1.5\arcsec\, $\sim$0.5\arcsec\ and
$\sim$0.18\arcsec. The contour maps for each source are all of equal
size and all contours increase by a factor of $\sqrt{2}$. \protect\label{comp_im_h}}
\end{figure*}
\begin{figure*}
\hspace{5mm}
\begin{minipage}{15cm}
\includegraphics[scale=0.85, angle=0]{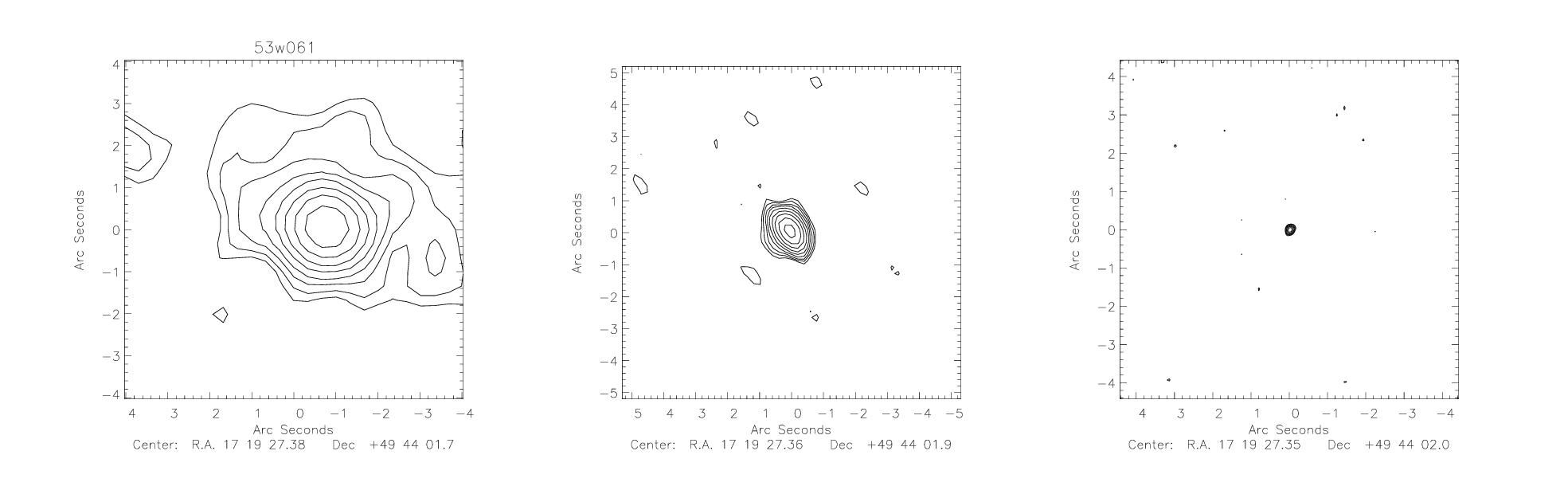}
\includegraphics[scale=0.85, angle=0]{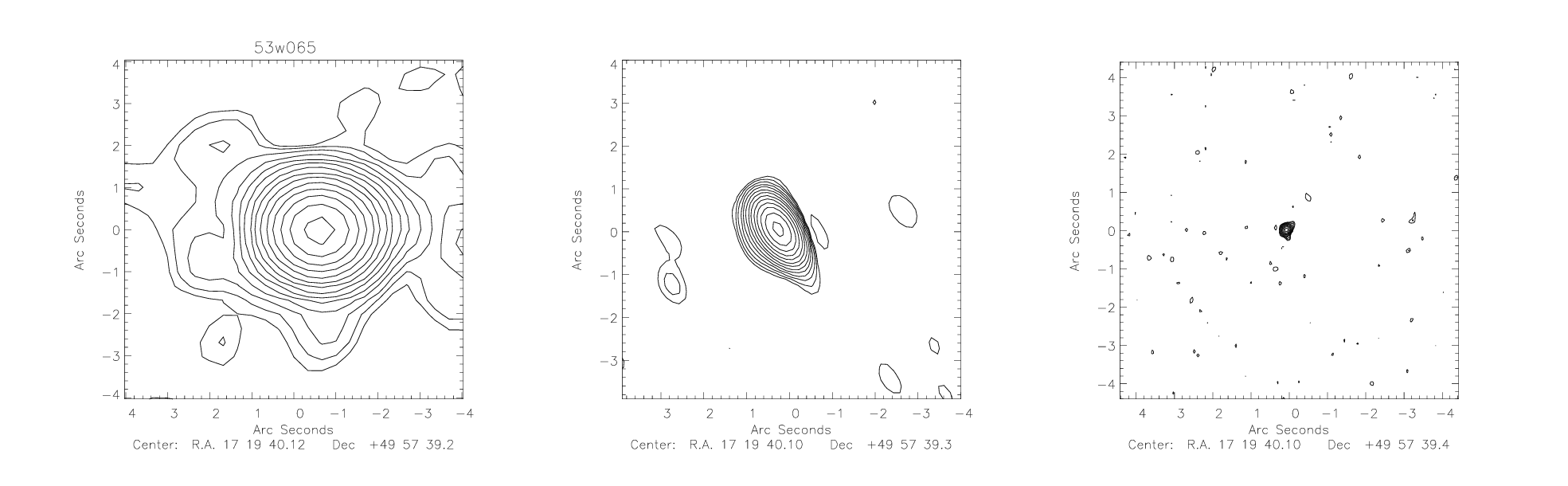}
\includegraphics[scale=0.85, angle=0]{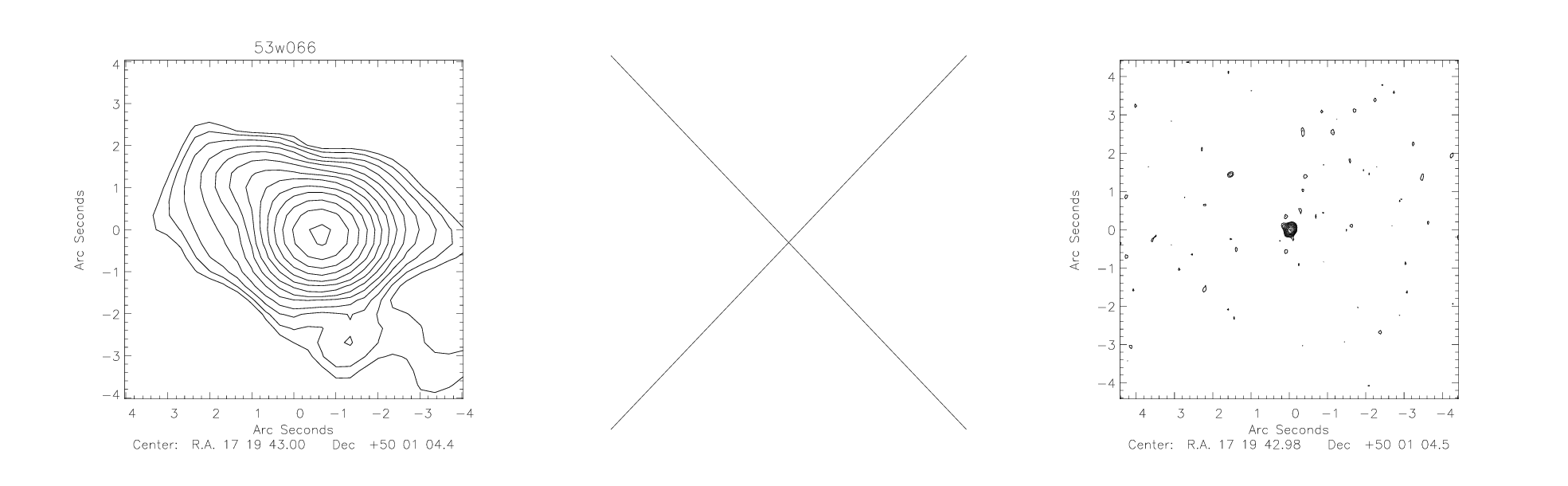}
\includegraphics[scale=0.85, angle=0]{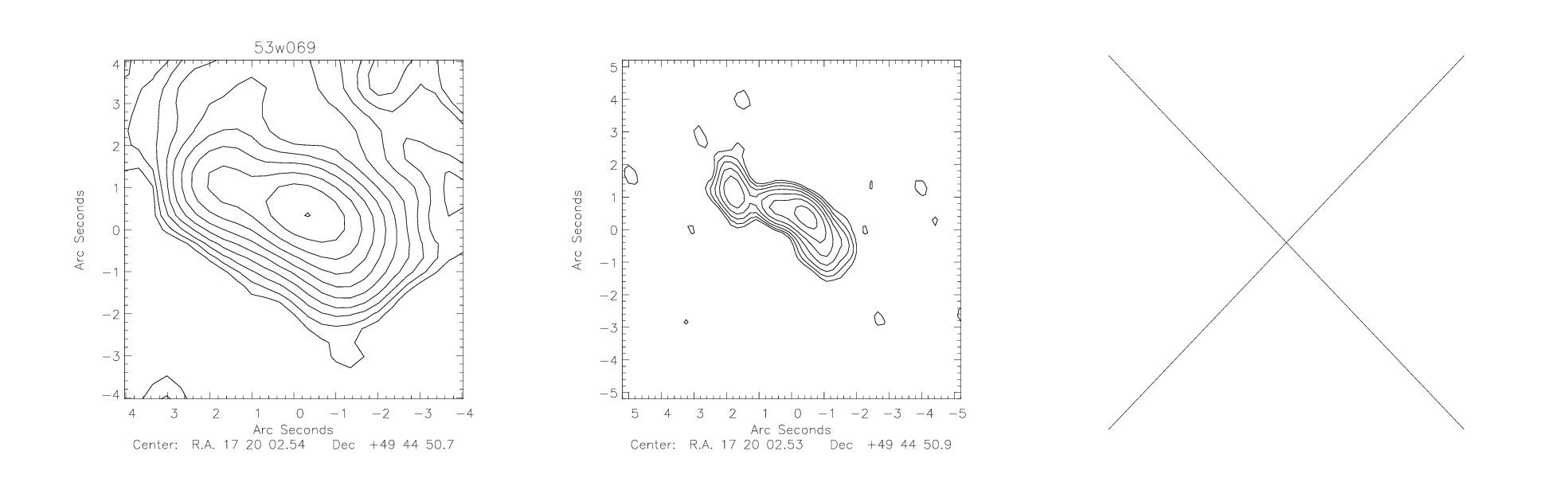}
\end{minipage}
\contcaption{}
\end{figure*}
\begin{figure*}
\hspace{5mm}
\begin{minipage}{15cm}
\includegraphics[scale=0.85, angle=0]{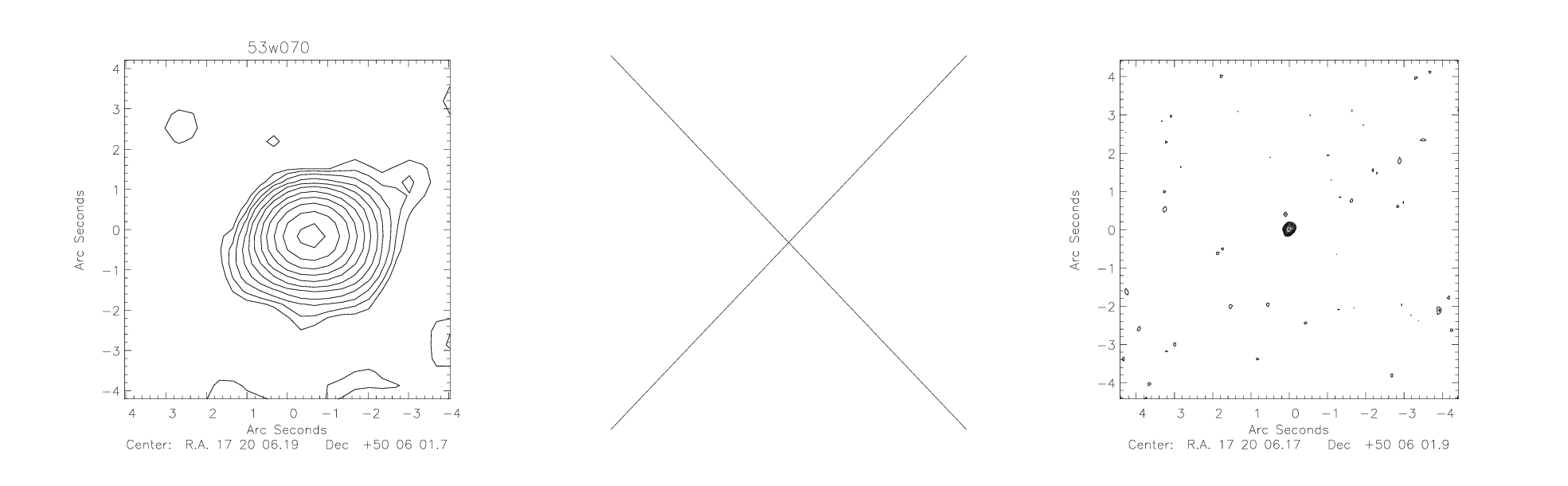}
\includegraphics[scale=0.85, angle=0]{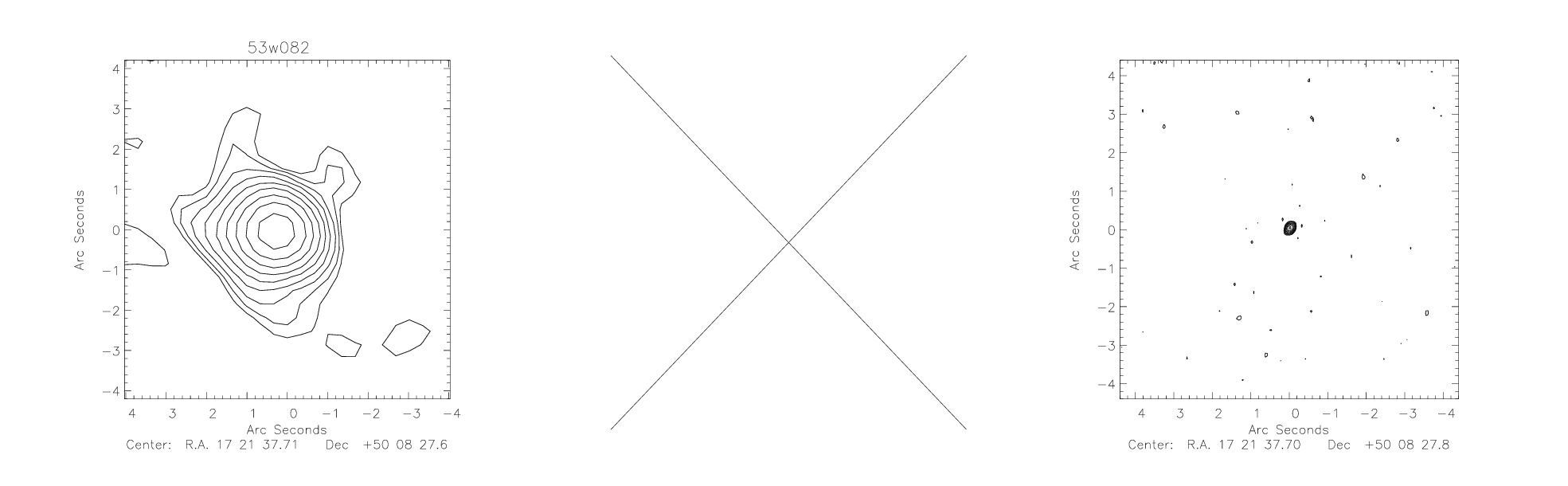}
\includegraphics[scale=0.85, angle=0]{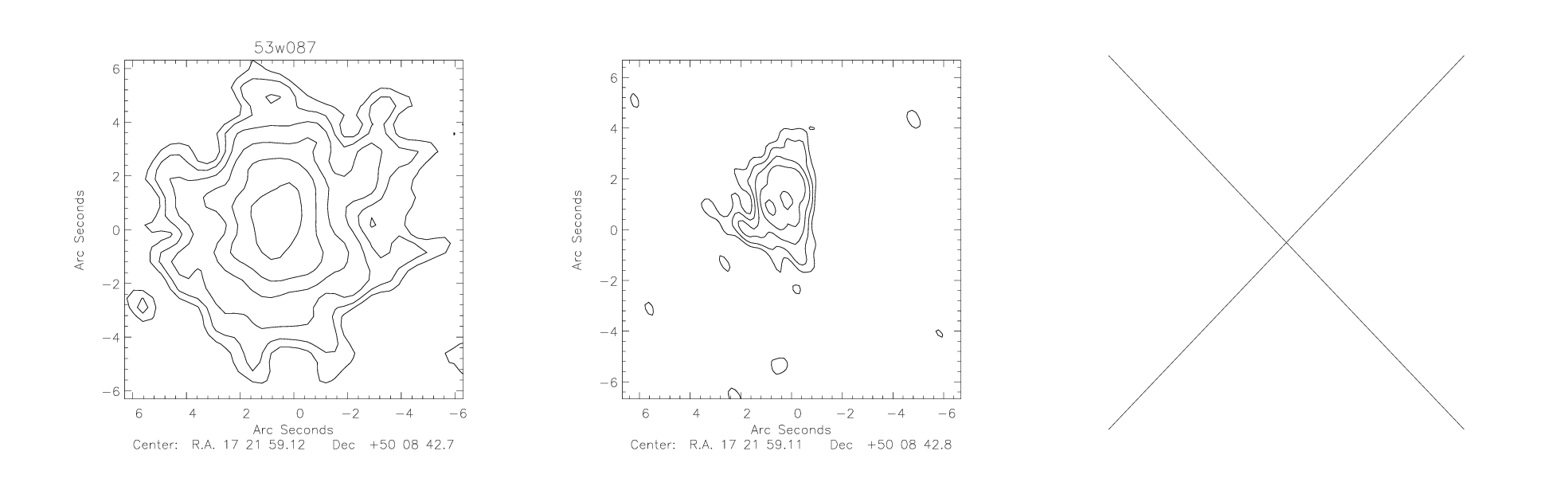}
\includegraphics[scale=0.85, angle=0]{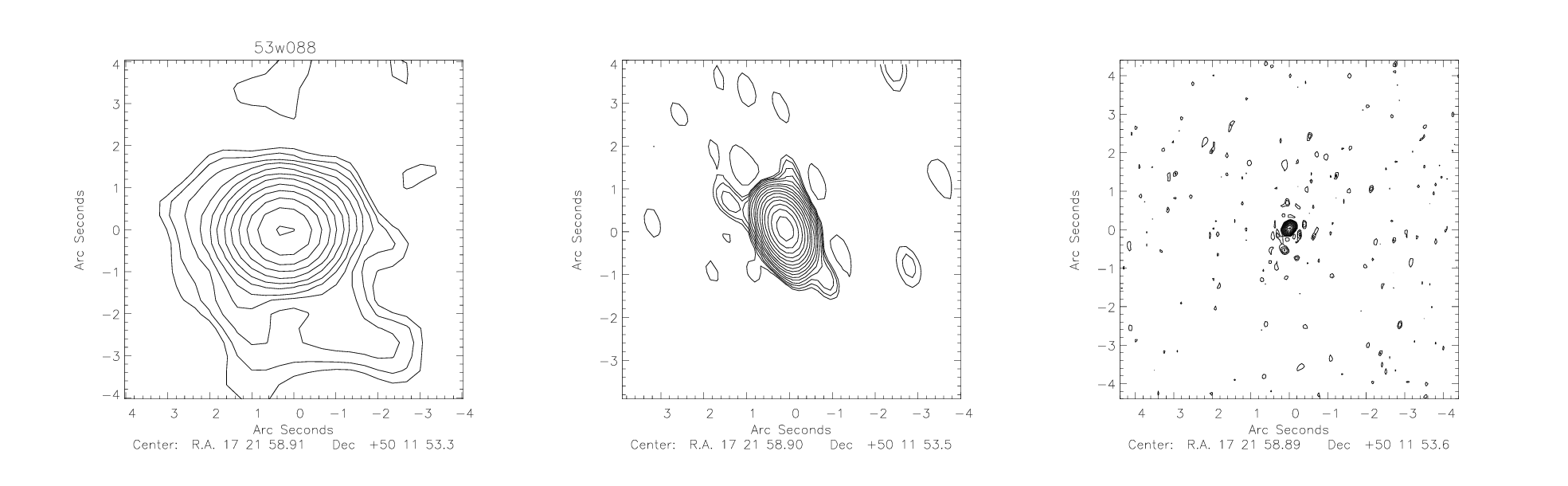}
\end{minipage}
\contcaption{}
\end{figure*}

\begin{figure*}
\centering
\begin{minipage}{20cm}
\includegraphics[scale=0.85, angle=90]{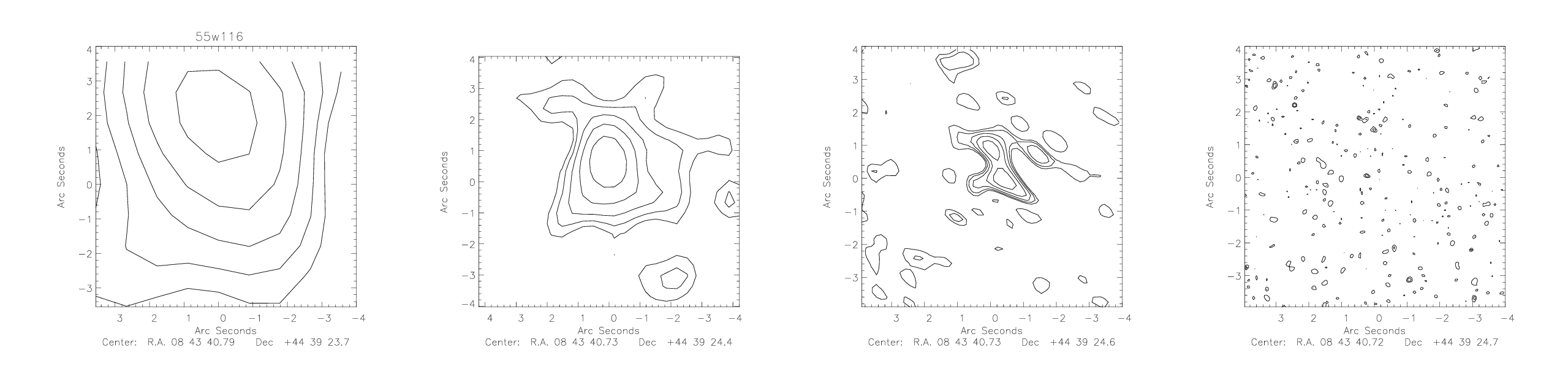}
\includegraphics[scale=0.85, angle=90]{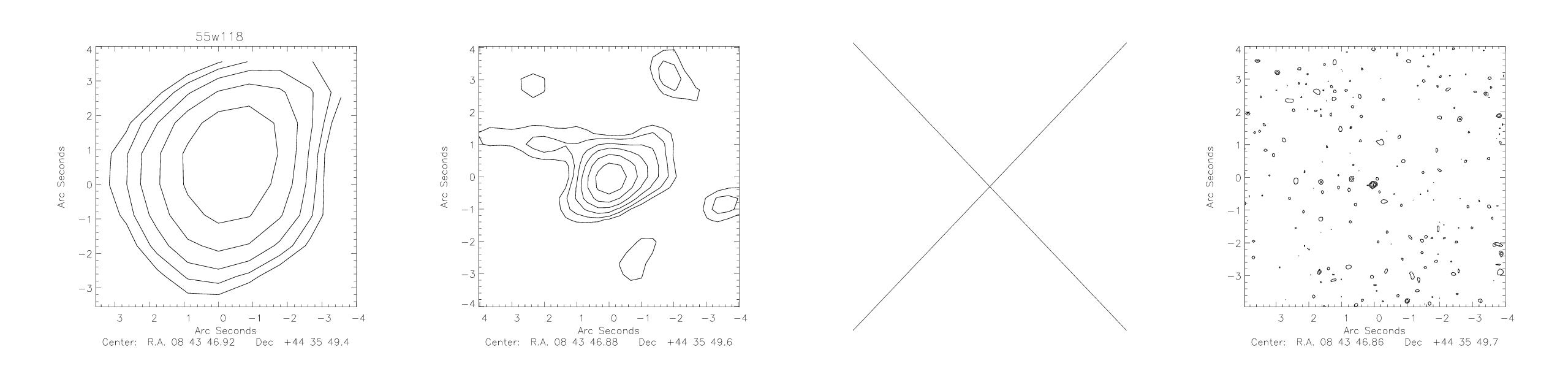}
\includegraphics[scale=0.85, angle=90]{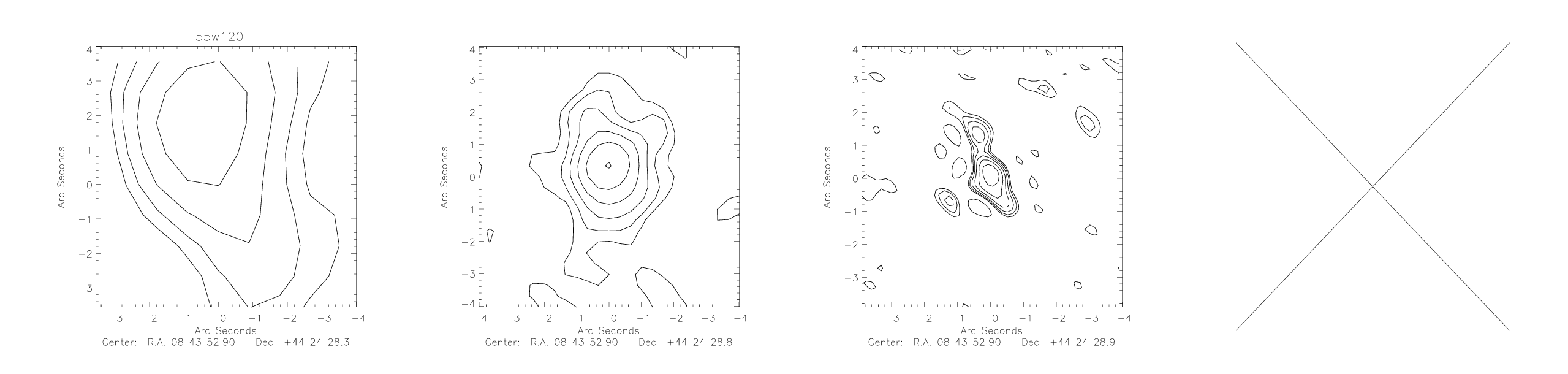}
\end{minipage}
\caption{({\it landscape}) The B--array, A--array, A+Pt and MERLIN contour maps for the sources in
the Lynx field, included in the A+Pt or MERLIN observations, all
centred on the optical host galaxy position if 
available, with the aim of comparing the source morphologies at the four different
resolutions, $\sim$5\arcsec,  $\sim$1.5\arcsec, $\sim$0.5\arcsec and
$\sim$0.18\arcsec. The contour maps for each source are all of equal
size and all contours increase by a factor of $\sqrt{2}$. \protect\label{comp_im}}
\end{figure*}
\begin{figure*}
\begin{minipage}{20cm}
\includegraphics[scale=0.85, angle=90]{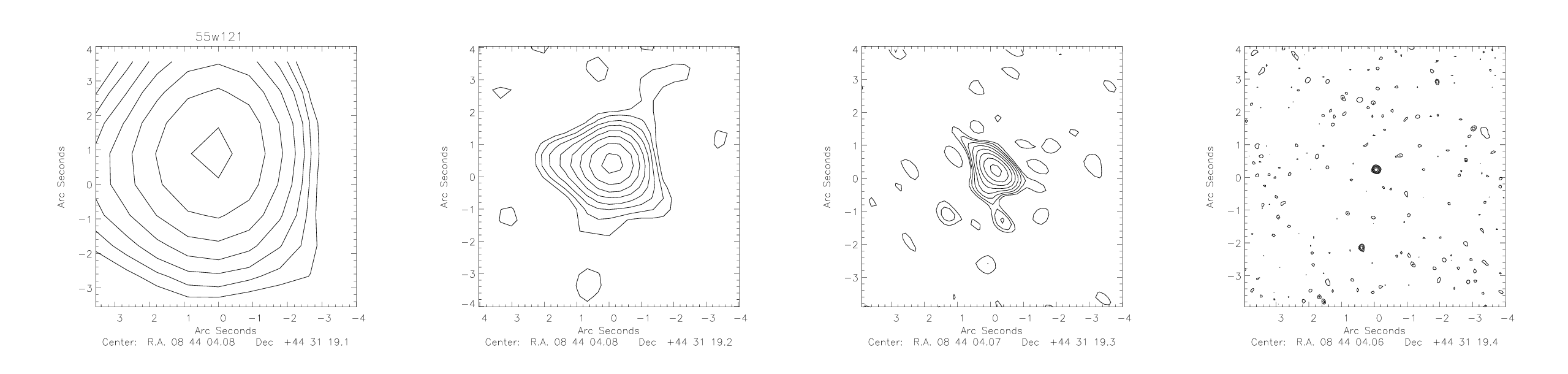}
\includegraphics[scale=0.85, angle=90]{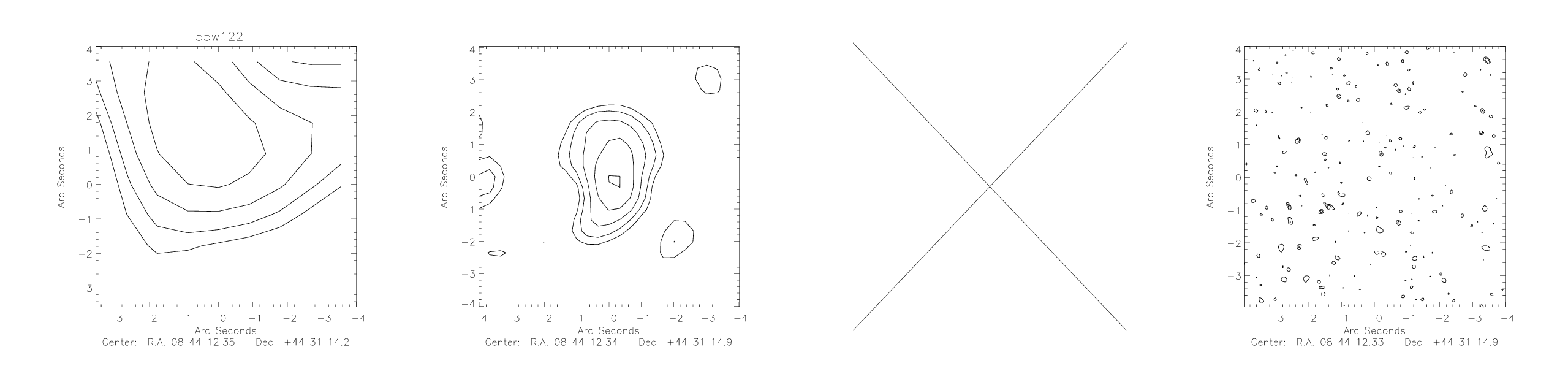}
\includegraphics[scale=0.85, angle=90]{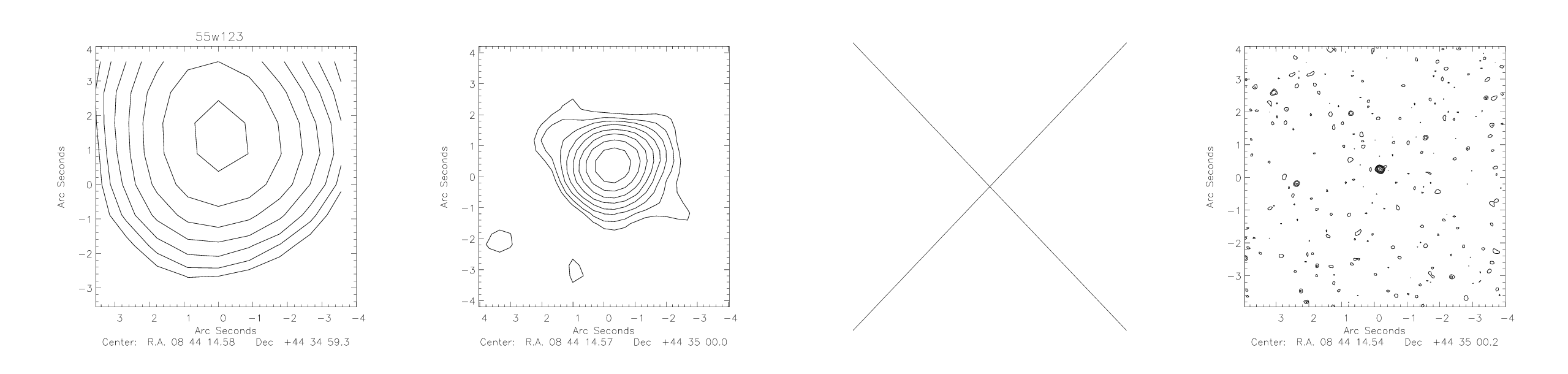}
\end{minipage}
\contcaption{}
\end{figure*}
\begin{figure*}
\begin{minipage}{20cm}
\includegraphics[scale=0.85, angle=90]{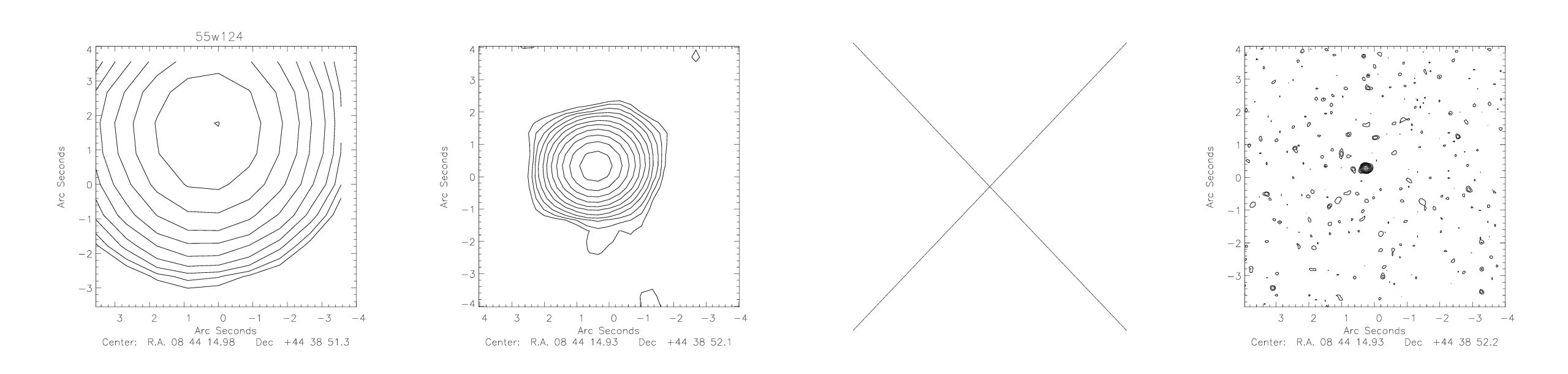}
\includegraphics[scale=0.85, angle=90]{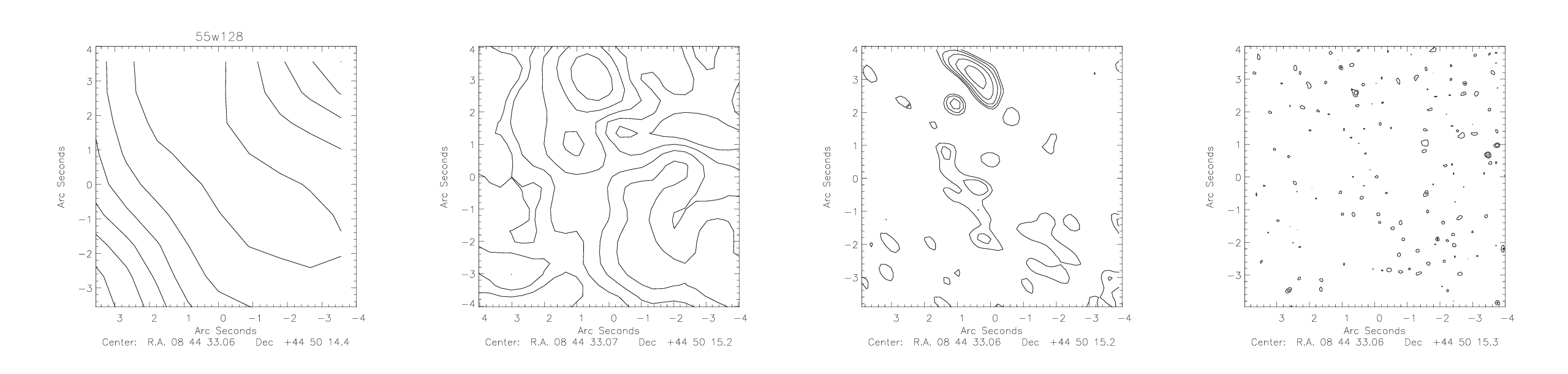}
\includegraphics[scale=0.85, angle=90]{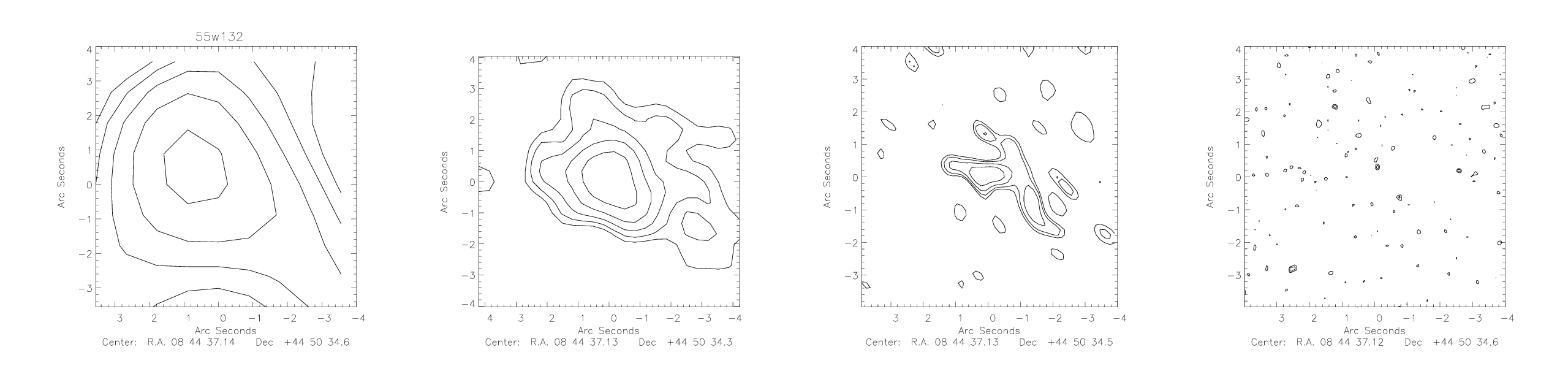}
\end{minipage}
\contcaption{}
\end{figure*}
\begin{figure*}
\begin{minipage}{20cm}
\includegraphics[scale=0.85, angle=90]{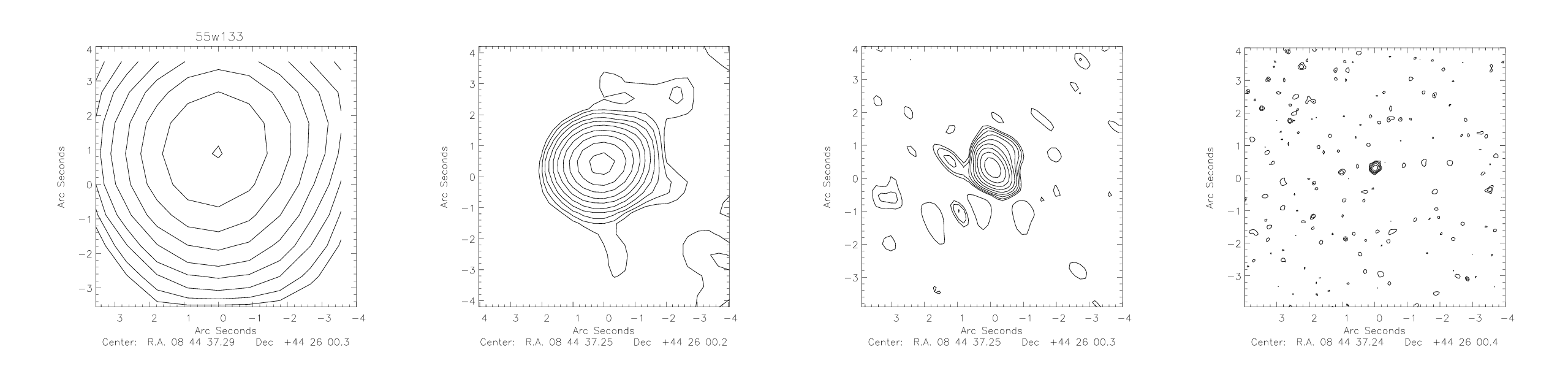}
\includegraphics[scale=0.85, angle=90]{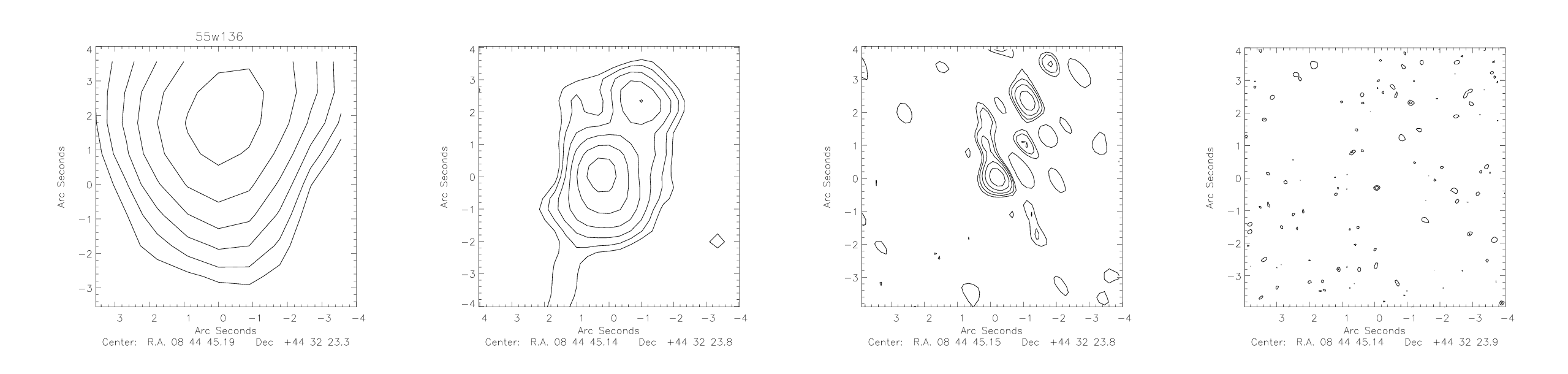}
\includegraphics[scale=0.85, angle=90]{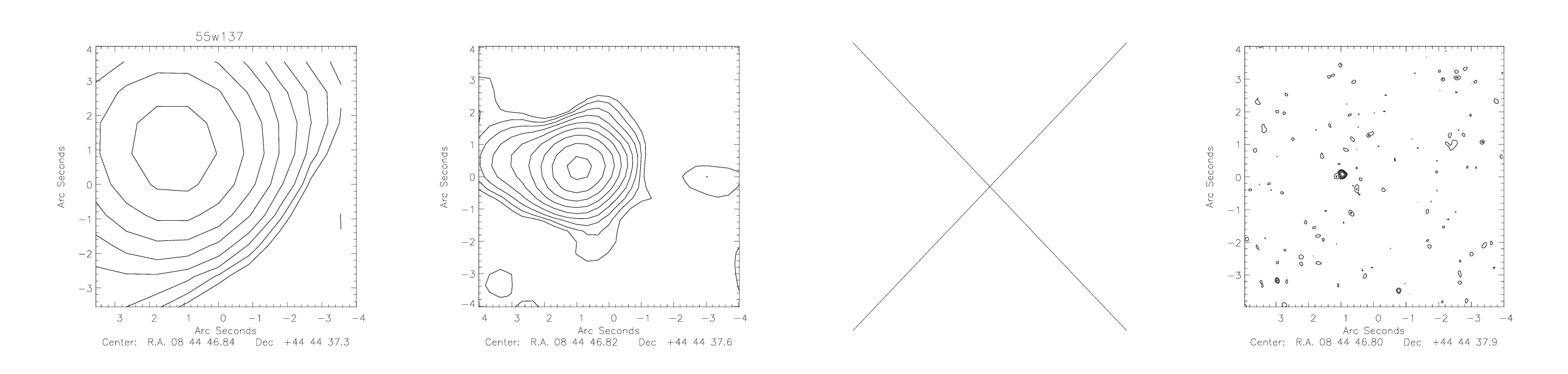}
\end{minipage}
\contcaption{}
\end{figure*}
\begin{figure*}
\begin{minipage}{20cm}
\includegraphics[scale=0.85, angle=90]{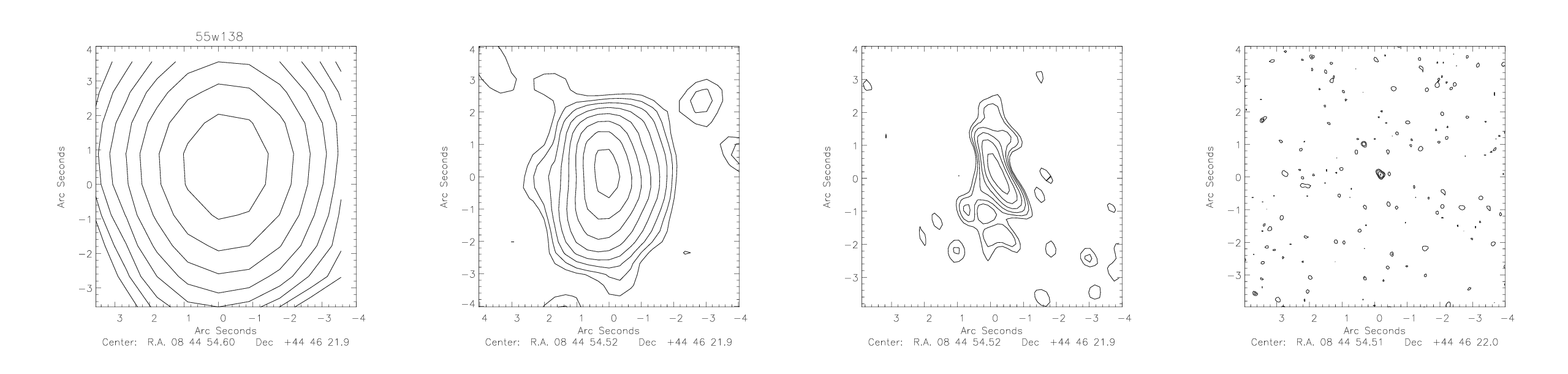}
\includegraphics[scale=0.85, angle=90]{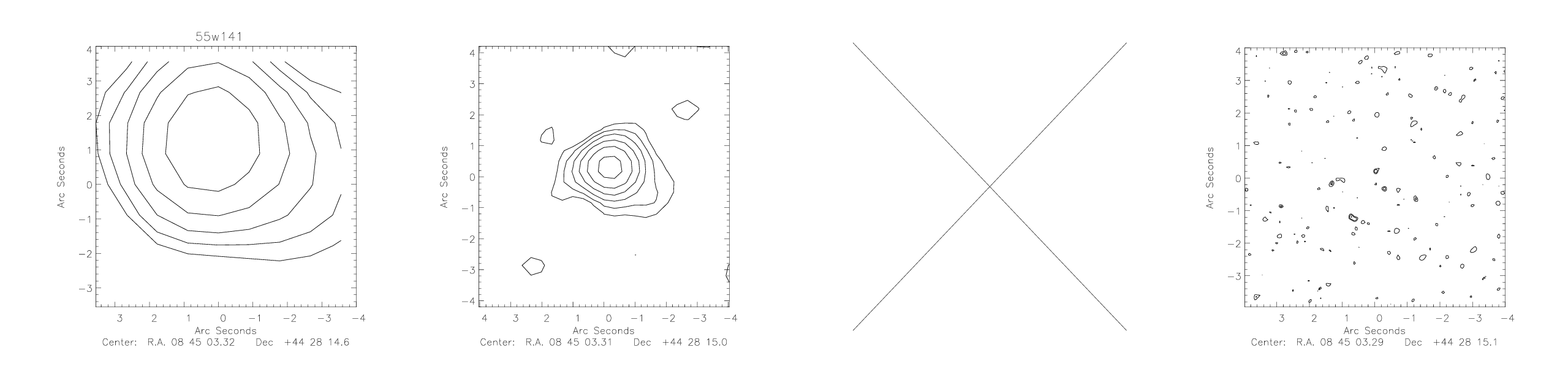}
\includegraphics[scale=0.85, angle=90]{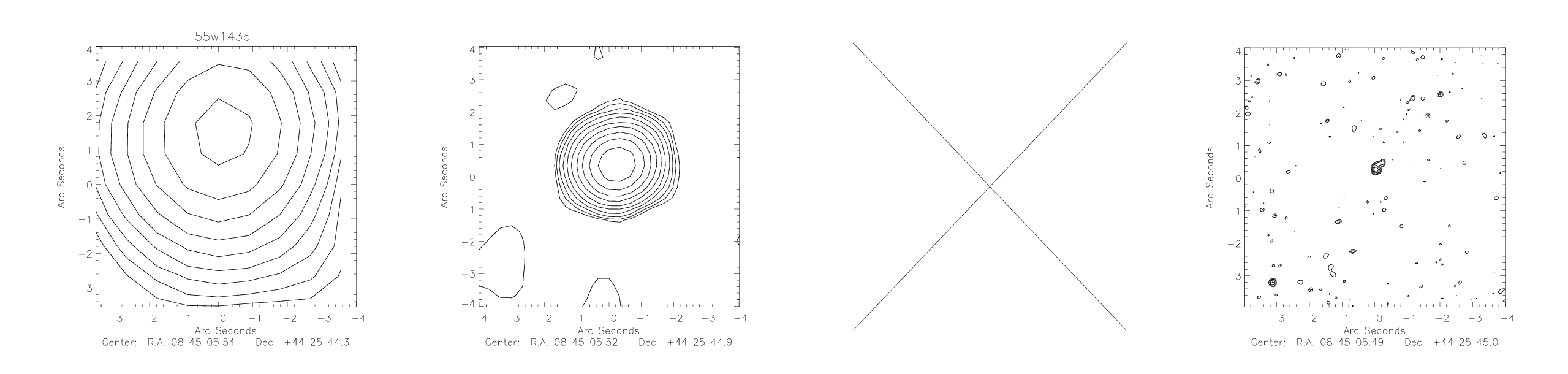}
\end{minipage}
\contcaption{}
\end{figure*}
\begin{figure*}
\begin{minipage}{20cm}
\includegraphics[scale=0.85, angle=90]{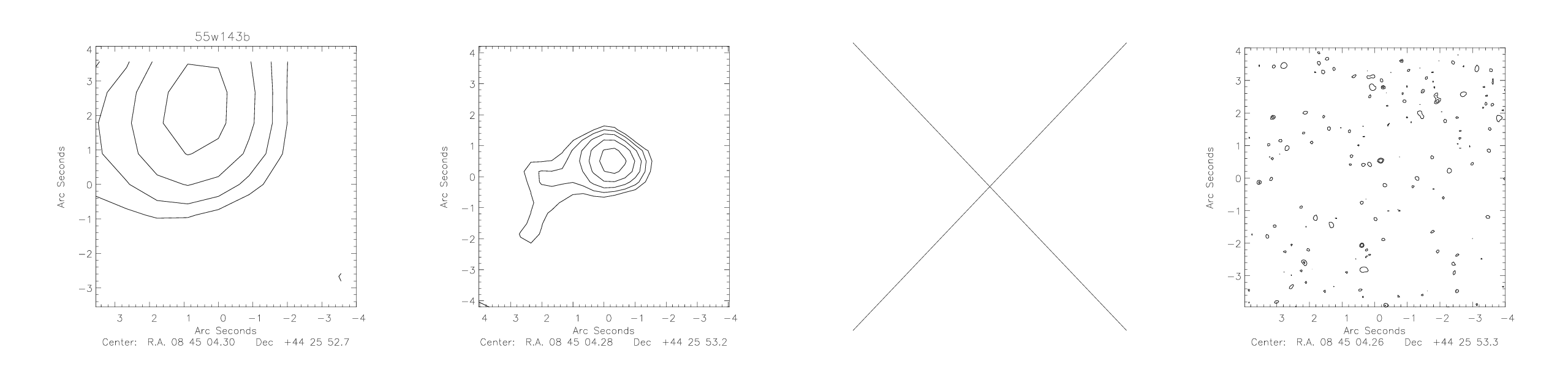}
\includegraphics[scale=0.85, angle=90]{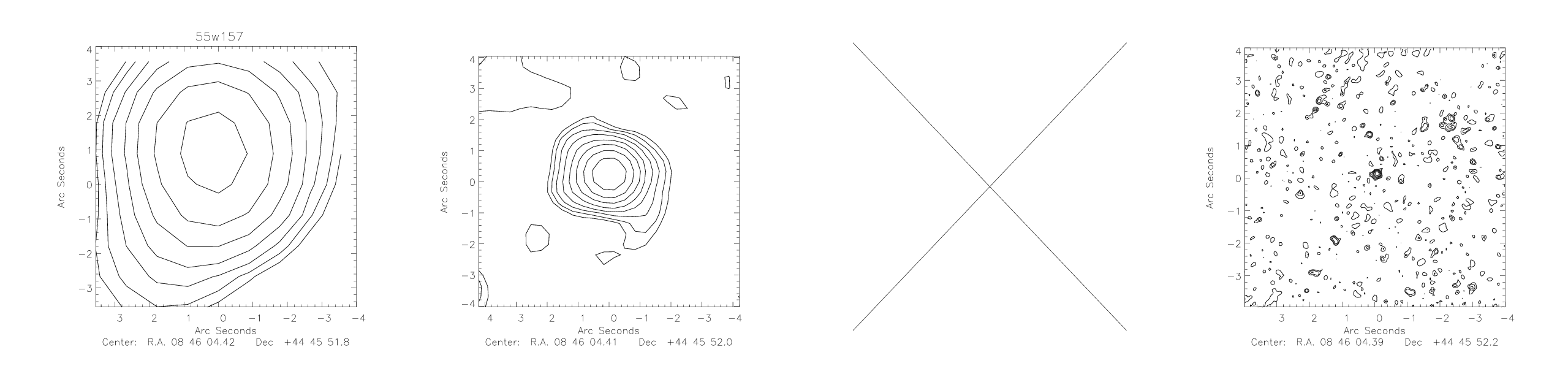}
\includegraphics[scale=0.85, angle=90]{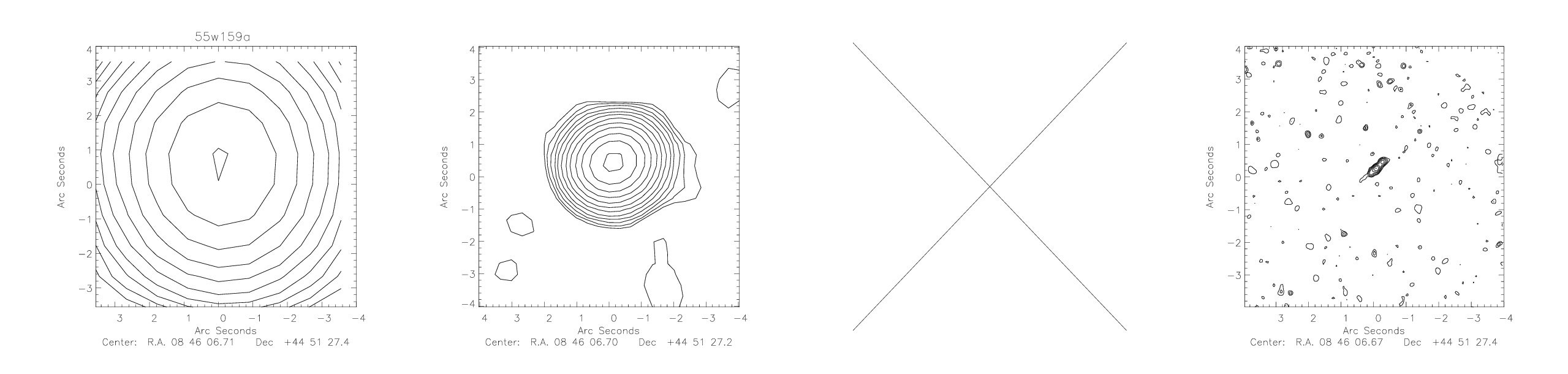}
\end{minipage}
\contcaption{}
\end{figure*}
\begin{figure*}
\begin{minipage}{20cm}
\includegraphics[scale=0.85, angle=90]{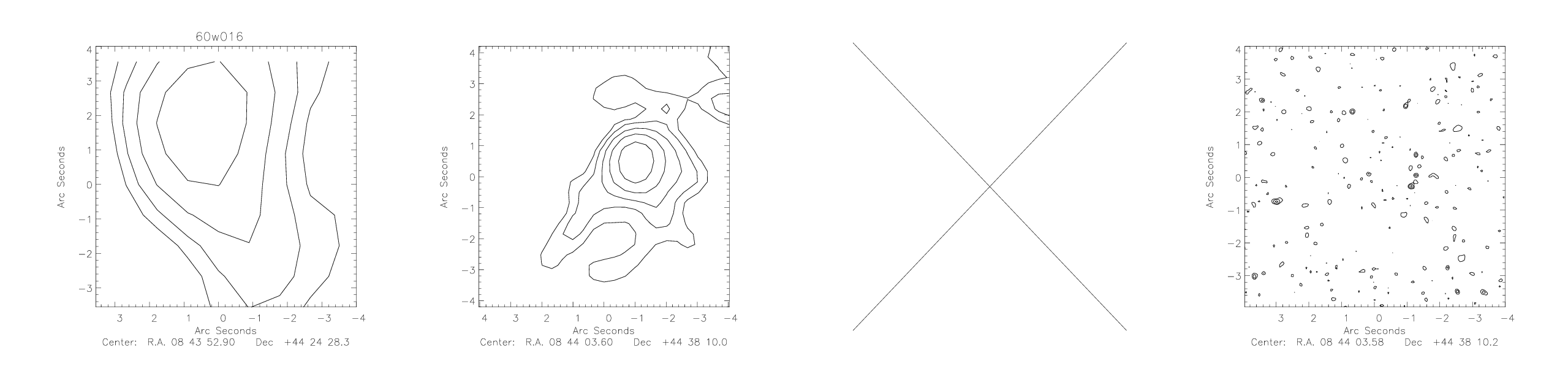}
\end{minipage}
\contcaption{}
\end{figure*}

\label{lastpage}

\end{document}